\documentclass[11pt]{article}

\usepackage{geometry}
\usepackage{booktabs}
\usepackage{diagbox}
\usepackage{cite}
\usepackage{enumerate}
\usepackage{graphicx}
\usepackage{subfigure}
\usepackage{hyperref}
\usepackage{amssymb}
\usepackage{amsmath}
\allowdisplaybreaks[1]

\usepackage{amsthm}

\newtheorem{theorem}{Theorem}
\newtheorem{lemma}[theorem]{Lemma}

\newtheorem{claim}[theorem]{Claim}
\newtheorem{observation}[theorem]{Observation}

\makeatletter
\def\@endtheorem{\endtrivlist}
\makeatother

\newcounter{rrule}
\renewcommand{\therrule}{R\arabic{rrule}}
\newenvironment{rrule}{\refstepcounter{rrule}\par\smallskip\noindent
\textbf{(\therrule)}\quad}{}

\newcounter{brule}
\renewcommand{\thebrule}{B\arabic{brule}}
\newenvironment{brule}{\refstepcounter{brule}\par\smallskip\noindent
\textbf{(\thebrule)}\quad}{}

\newcommand{\instance}[2]{#1[#2]}
\newcommand{\add}[2]{\mathrm{Added}(#1,#2)}
\newcommand{\rem}[2]{\mathrm{Removed}(#1,#2)}

\newcommand{\hsb}{\textsc{HS}}
\newcommand{\hs}[1]{\hsb(#1)}

\begin{document}

\title{Faster parameterized algorithm for \textsc{$3$-Hitting Set}}
\author{Dekel Tsur%
\thanks{Department of computer science, Ben-Gurion University of the Negev.
Email: \texttt{dekelts@cs.bgu.ac.il}}}
\date{}
\maketitle

\begin{abstract}
In the \textsc{$3$-Hitting Set} problem,
the input is a hypergraph $G$ such that the size of every hyperedge of $G$ is at most 3,
and an integers $k$,
and the goal is to decide whether there is a set $S$ of at most $k$ vertices
such that every hyperedge of $G$ contains at least one vertex from $S$.
In this paper we give an $O^*(2.0409^k)$-time algorithm for \textsc{$3$-Hitting Set}.
\end{abstract}


\section{Introduction}
In the \textsc{$d$-Hitting Set} problem,
the input is a hypergraph $G$ such that the size of every hyperedge of $G$ is at most $d$,
and an integers $k$,
and the goal is to decide whether there is a set $S$ of at most $k$ vertices
such that every hyperedge of $G$ contains at least one vertex from $S$.
Many problem can be reduced to \textsc{$d$-Hitting Set} for some fixed $d$.
For example, \textsc{Maximum Agreement Subtree} and
\textsc{Call Control} in undirected trees of rings can be reduced to
\textsc{$d$-Hitting Set}~\cite{anand2003call,downey1999parameterized}.
Additionally, every vertex deletion problem into $\mathcal{H}$-free graph,
where $\mathcal{H}$ is a finite set of forbidden graph,
can be reduced to \textsc{$d$-Hitting Set}.

The \textsc{$2$-Hitting Set} problem is also called \textsc{Vertex Cover}.
For the \textsc{Vertex Cover} problem,
Downey and Fellows~\cite{downey1995parameterized} gave an $O^*(2^k)$-time algorithm,
Balasubramanian et al.~\cite{balasubramanian1998improved} gave an $O^*(1.325^k)$-time algorithm,
Downey et al.~\cite{downey1999parameterized} gave an $O^*(1.320^k)$-time algorithm,
Niedermeier and Rossmanith~\cite{niedermeier1999upper} gave an $O^*(1.292^k)$-time algorithm,
Stege and Fellows~\cite{stege1999improved} gave an $O^*(1.291^k)$-time algorithm,
Chen et al.~\cite{chen2001vertex} gave an $O^*(1.286^k)$-time algorithm,
Niedermeier and Rossmanith~\cite{niedermeier2003efficient2} gave an $O^*(1.284^k)$-time algorithm
(with exponential space),
Chandran and Grandoni~\cite{chandran2004refined} gave an $O^*(1.275^k)$-time algorithm
(with exponential space),
Chen et al.~\cite{chen2010improved} gave an $O^*(1.274^k)$-time algorithm, and
Harris and Narayanaswamy~\cite{harris2022faster} gave an $O^*(1.254^k)$-time algorithm.

For the \textsc{$3$-Hitting Set} problem,
Bryant et al.~\cite{bryant1998parameterized} gave an $O^*(2.562^k)$-time algorithm,
Niedermeier and Rossmanith~\cite{niedermeier2003efficient} gave an $O^*(2.270^k)$-time algorithm,
Fernau~\cite{fernau2010top} gave an $O^*(2.179^k)$-time algorithm, and
Wahlstr{\"o}m~\cite{wahlstrom2007algorithms} gave an $O^*(2.0755^k)$-time algorithm.

For larger values of $d$,
Niedermeier and Rossmanith~\cite{niedermeier2003efficient} gave an
algorithm for \textsc{$d$-Hitting Set} whose running time is $O^*(c_d^k)$,
where $c_d = \frac{d-1}{2}(1+\sqrt{1+4/(d-1)^2})$.
The values of $c_d$ for $d=4,5,6,7$ are approximately 3.303, 4.237, 5.193, and 6.163, respectively.
Fernau~\cite{fernau2010parameterized} gave an algorithm for \textsc{$d$-Hitting Set}
with smaller exponential bases.
The exponential bases of Fernau's algorithm for $d=4,5,6,7$ are approximately
3.115, 4.08, 5.049, and 6.033, respectively.
Dom et al.~\cite{dom2010fixed} and Fomin et al.~\cite{fomin2010iterative} showed that
if there is an $O^*(c^k)$-time algorithm for \textsc{$d$-Hitting Set},
then there is an $O^*((c+1)^k)$-time algorithm for \textsc{$(d+1)$-Hitting Set}.
Combining this result with the algorithm of Wahlstr{\"o}m~\cite{wahlstrom2007algorithms}
for \textsc{$3$-Hitting Set} yields an $O^*((d-1+0.0755)^k)$-time algorithm for
\textsc{$d$-Hitting Set}.
This algorithm is faster than the algorithm of Fernau~\cite{fernau2010top} when $d \leq 5$.
Kamat et al.~\cite{kamat2015parameterized} and Gupta et al.~\cite{gupta2021parameterized}
studied the \textsc{$d$-Hitting Set} problem with additional constraints on
the sizes of the intersections between the hitting set and the hyperedges.

In this paper we give an $O^*(2.0409^k)$-time algorithm for \textsc{$3$-Hitting Set}.
This result yields an $O^*((d-1+0.0409)^k)$-time algorithm for \textsc{$d$-Hitting Set},
which is faster than the algorithm of Fernau~\cite{fernau2010top} when $d \leq 6$.
Our algorithm is based on the algorithm of Wahlstr{\"o}m~\cite{wahlstrom2007algorithms}.
Our algorithm uses the reduction and branching rules from~\cite{wahlstrom2007algorithms}.
Some of these rules were modified by changing the choices of the vertices on which the rules are applied.
Additionally, our algorithm uses new reduction and branching rules
(Rules~(\ref{rule:cc}), (\ref{rule:deg2-u}), (\ref{rule:cycle}), (\ref{rule:deg2-33-pre1}),
and (\ref{rule:deg2-33-pre2})).
The better running time is achieved by using a more involved analysis.
Namely, when analyzing some branching rules of the algorithm,
we consider the structures of the hypergraphs in the instances generated by the rules
and show that the algorithm applies ``good'' rules on these instances.
Additionally, our analysis uses a measure function that is different than
the one in~\cite{wahlstrom2007algorithms}.

\section{Preliminaries}
For integers $a \leq b$, $[a,b] = \{a,a+1,\ldots,b\}$.
For integers $a,b,c$, $[a,b]_c$ is equal to $[a,b]$ if $a \leq b$
and otherwise $[a,b]_c = \{c\}$.

A \emph{hypergraph} is a pair $(V,E)$, where $V$ is some set of elements
and $E$ is a set whose elements are nonempty subsets of $V$.
An element of $V$ is called a \emph{vertex} and an element of $E$ is called
a \emph{hyperedge}.
For a hypergraph $G$, $V(G)$ is the set of the vertices of $G$ and
$E(G)$ is the set of the hyperedges of $G$.
A hyperedge of size $s$ is called an \emph{$s$-hyperedge}.
A \emph{3-hypergraph} is a hypergraph in which every hyperedge has size at most 3.

Let $G$ be a hypergraph and let $v$ be a vertex in $G$.
A vertex $w$ is a \emph{neighbor} of $v$ if 
there is at least one hyperedge that contains both $v$ and $w$.
The set of all neighbors of $v$ is denoted $N^G(v)$.
The \emph{degree} of $v$, denoted $d^G(v)$, is the number of hyperedges that contain $v$.
Let $d_s^G(v)$ be the number of $s$-hyperedges that contain $v$.
Let $D_2^G(v)$ be the number of 2-hyperedges $e$ such that either
(1) $v \in e$, or (2) there is a 2-hyperedge $e'$ such that $v \in e'$ and $e \cap e' \neq \emptyset$.
Let $I^G(v) = 1$ if there is a hyperedge $e$ such that $v \notin e$ and $|e \cap N^G(v)| \geq 2$.
We will omit the superscript $G$ when the hypergraph $G$ is clear from the context.

For a hypergraph $G$,
$d(G) = \max_{v \in V(G)} d^G(v)$,
$d_s(G) = \max_{v \in V(G)} d_s^G(G)$,
$E_s(G)$ is the set of all $s$-hyperedges in $G$,
and $m_s(G) = |E_s(G)|$.
For a hypergraph $G$,
$G^2$ is the graph $(\bigcup_{e \in E_2(G)} e, E_2(G))$, and
$c_2(G)$ is the number of connected components of $G^2$.

For a hypergraph $G$ and a vertex $v$,
$\instance{G}{+v}$ is the hypergraph $(V(G) \setminus \{v\}, \{e : e \in E(G), v \notin e\})$
and
$\instance{G}{-v}$ is the hypergraph $(V(G) \setminus \{v\}, \{e \setminus \{v\} : e \in E(G)\})$.
Note that $(G,k)$ is a yes-instance if and only if
$(\instance{G}{+v},k-1)$ is a yes-instance or $(\instance{G}{-v},k)$ is a yes-instance
(assuming $(G,k)$ is a yes-instance,
$(\instance{G}{+v},k-1)$ is a yes-instance if there is a hitting set of $G$ of size at most $k$
that contains $v$, and
$(\instance{G}{-v},k)$ is a yes-instance if there is a hitting set of $G$ of size at most $k$
that does not contain $v$).
If $\alpha_i$ is either $+v_i$ or $-v_i$ for $i = 1,\ldots,s$, then
$\instance{G}{\alpha_1,\ldots,\alpha_s} = \instance{(\instance{G}{\alpha_1,\ldots,\alpha_{s-1}})}{\alpha_s}$.

A hypergraph $G$ is \emph{simple} if there are no hyperedges $e,e'$ of $G$ such that $e \subset e'$.
A vertex $u$ is \emph{dominated} by a vertex $v$ if every hyperedge that contains $u$ also contains $v$.

A \emph{branching algorithm} is a recursive algorithm such that given an instance $(G,k)$,
it either solves the instance or applies a \emph{reduction rule} or a \emph{branching rule}.
If the algorithm applies a reduction rule, then it generates an instance $(G',k')$ and
makes a recursive call on $(G',k')$.
If the algorithm applies a branching rule, then it generates instances $(G_1,k_1),\ldots,(G_s,k_s)$
and makes recursive calls on these instances.
A reduction rule is \emph{safe} if $(G,k)$ is a yes-instance if and only if
$(G',k')$ is a yes-instance.
A branching rule is \emph{safe} if $(G,k)$ is a yes-instance if and only if there is
an index $i \in [1,s]$ such that $(G_i,k_i)$ is a yes-instance.
In the \emph{measure and conquer approach}, a branching algorithm is analyzed using
a \emph{measure function} $\mu$ that maps instances $(G,k)$ to real numbers.
If a branching rule generates instances $(G_1,k_1),\ldots,(G_s,k_s)$ from an instance $(G,k)$,
then the vector $(\mu(G,k)-\mu(G_1,k_1),\ldots,\mu(G,k)-\mu(G_s,k_s))$ is called
a \emph{branching vector} of the rule.
Note that a branching rule can have several branching vectors.
The \emph{branching number} of a branching vector $(a_1,\ldots,a_s)$
is the maximum real solution of the equality $\sum_{i=1}^s x^{-a_i} = 1$.
The branching number of a branching rule is the supremum of the branching numbers of its branching vectors.
If all branching rules of an algorithm have branching numbers at most $\alpha$,
then the running time of the algorithm is $O^*(\alpha^{\mu(G,k)})$
(assuming each rule can be applied in polynomial time and that instances that are mapped by $\mu$
to negative values can be solved in polynomial time).

A vector $(a_1,\ldots,a_s)$ is \emph{dominated} by a vector $(b_1,\ldots,b_t)$ if there is
an injective mapping $f \colon \{1,\ldots,s\} \to \{1,\ldots,t\}$ such that
$a_i \geq b_{f(i)}$ for all $i$.
If a vector $X_1$ is dominated by a vector $X_2$, then the branching number of $X_1$
is less than or equal to the branching number of $X_2$.
A set of vectors $A$ is dominated by a set of vectors $B$ if
every vector in $A$ is dominated by a vector in $B$.

\section{The algorithm}

We now describe an algorithm \hsb\ for solving \textsc{$3$-Hitting Set}.
If $k < 0$, then the algorithm returns \textsc{False}.
If $E(G) = \emptyset$, then the algorithm returns \textsc{True}.
Otherwise, the algorithm applies the first applicable rule from the
rules below.

\begin{rrule}\label{rule:loop}
If $\{x\}$ is a hyperedge in $G$, then return $\hs{\instance{G}{+x},k-1}$.
\end{rrule}

\begin{rrule}\label{rule:non-simple}
If $G$ is not simple, then return $\hs{G',k}$,
where $G'$ is a hypergraph that contains all the minimal hyperedges of $G$.
\end{rrule}

\begin{rrule}\label{rule:cc}
If there is a connected component $H$ of $G$ with at least two vertices
that has a hitting set of size at most 3,
then compute a minimum size hitting set $S_H$ of $H$ and return $\hs{G-H,k-|S_H|}$.
\end{rrule}

\begin{rrule}\label{rule:dominated}
If $x$ is a vertex that is dominated by another vertex, then return $\hs{\instance{G}{-x},k}$.
\end{rrule}

\begin{rrule}\label{rule:deg2-c}
If there is a vertex $x$ with degree~2 and there is a hyperedge $e$ such that
$e \subseteq N(x)$, then return $\hs{\instance{G}{-x},k}$.
\end{rrule}

\begin{rrule}\label{rule:deg2-u}
If there are vertices $x,u$ such that $u \in e$ for every hyperedge $e$ such that
$e \cap N(x) \neq \emptyset$ and $x \notin e$, then
return $\hs{\instance{G}{+x},k-1}$.
\end{rrule}

\begin{brule}\label{rule:cycle}
If $d_2(G) = 2$ and there are vertices $v_1,\ldots,v_4$ and hyperedges $\{v_1,v_2\}$, $\{v_2,v_3\}$,
$\{v_3,v_4\}$, and $\{v_1,v_4\}$, then return
$\hs{\instance{G}{+v_1,+v_3},k-2} \lor \hs{\instance{G}{+v_2,+v_4},k-2}$.
\end{brule}

\begin{brule}\label{rule:deg2-2x}
If $d_2(G) \in \{1,2\}$ and there is a vertex $x$ such that $d_2(x) = d_2(G)$
and $d(x) = 2$, then
\begin{enumerate}
\item
If $d_2(G) = 2$, then let $e_1 = \{x,y\}$ and $e_2 = \{x,z\}$ be the hyperedges containing $x$.
Return $\hs{\instance{G}{+x,-y,-z},k-1} \lor \hs{\instance{G}{-x,+y,+z},k-2}$.
\item 
If $d_2(G) = 1$, then let $e_1 = \{x,y\}$ and $e_2 = \{x,z,w\}$ be the hyperedges containing $x$.
Return $\hs{\instance{G}{+x,-y,-z,-w},k-1} \lor \hs{\instance{G}{-x,+y},k-1}$.
\end{enumerate}
\end{brule}

\begin{brule}\label{rule:d2}
If $m_2(G) > 0$, then let $x$ be a vertex such that
$d_2(x) = d_2(G)$ and
$d_3(x)-D_2(x) \geq d_3(x')-D_2(x')$ for every vertex $x'$ such that $d_2(x') = d_2(G)$.
Return $\hs{\instance{G}{+x},k-1} \lor \hs{\instance{G}{-x},k}$.
\end{brule}

\begin{brule}\label{rule:deg2-33-pre1}
If $d(G) = 4$ and $G$ contains vertices $x,y$ and hyperedges $e_1,e_2,e_3$ such that
$x,y \in e_i$ for every $i$, then
return $\hs{\instance{G}{+x,-y},k-1} \lor \hs{\instance{G}{-x,+y},k-1} \lor \hs{\instance{G}{-x,-y},k}$.
\end{brule}

\begin{brule}\label{rule:deg2-33-pre2}
If $d(G) \leq 4$ and $G$ contains a vertex $x$ such that $d(x) = 2$, $I(x) = 0$,
$d(p) = 3$ for every $p \in N(x)$, and
$|B| = 4$, where $B = \{e \setminus N(x) : e \in E(G), x \notin e, e \cap N(x) \neq \emptyset\}$
(note that $B$ is a set and not a multi-set),
then let $H = (N(x),B,E_H)$ be a bipartite graph
in which there is an edge $\{p,e\}$ for $p \in N(x)$ and $e \in B$ if and only if
$\{p\} \cup e \in E(G)$.
Let $e_1,e_2$ be the hyperedges that contain $x$.
\begin{enumerate}
\item
If $H$ is a disjoint union of two chordless cycles of length~4, and
every cycle contains exactly one vertex from $e_1 \setminus \{x\}$
and exactly one vertex from $e_2 \setminus \{x\}$, then
let $e_1 = \{x,y,z\}$ and $e_2 = \{x,v,w\}$, where $y,v$ are in the same cycle in $H$, and
return $\hs{\instance{G}{+x,-y,-z,-v,-w},k-1} \lor \hs{\instance{G}{+y,+v,-x},k-2} \lor
\hs{\instance{G}{+z,+w,-x},\allowbreak k-2}$.
\item
Otherwise, let $e_1 = \{x,y,z\}$ and $e_2 = \{x,v,w\}$, and
return $\hs{\instance{G}{+x,-y,-z,-v,-w},\allowbreak k-1} \lor \hs{\instance{G}{+y,+z,+v,+w,-x},k-4}$.
\end{enumerate}
\end{brule}

\begin{brule}\label{rule:deg2-33}
If $d(G) \leq 4$ and $G$ contains vertices with degree~2, then let $x$ be a vertex such that
$d(x) = 2$ and
$I(x) \geq I(x')$ for every $x'$ such that $d(x') = 2$.
Let $e_1 = \{x,y,z\}$ and $e_2 = \{x,v,w\}$ be the hyperedges containing $x$,
and return $\hs{\instance{G}{+x,-y,-z,-v,-w},\allowbreak k-1} \lor \hs{\instance{G}{-x},k}$.
\end{brule}

\begin{brule}\label{rule:not-connected}
If $d(G) \leq 3$ and $G$ is not connected, then let $H$ be a connected component of $G$.
Perform recursive calls $\hs{H,k'}$ for $k' = 4,5,\ldots,k-4$ until
one call returns \textsc{True} or until the loop ends.
In the former case return $\hs{G-H,k-k'}$ and in the latter case return \textsc{False}.
\end{brule}

\begin{brule}\label{rule:final}
Let $u$ be a vertex such that $d(u) = d(G)$ and
$d_2(\instance{G}{-u}) \geq d_2(\instance{G}{-u'})$ for every vertex $u'$ such that $d(u') = d(G)$.
Return $\hs{\instance{G}{+u},k-1} \lor \hs{\instance{G}{-u},k}$.
\end{brule}

We can assume that the input $(G,k)$ to the algorithm satisfies that $G$
does not contain an empty hyperedge (otherwise $(G,k)$ is a no-instance).
The following lemma shows that empty hyperedges are not created during the algorithm.
\begin{lemma}\label{lem:empty}
Let $(G,k)$ be an instance and let $(G',k')$ be an instance that is obtained by
applying a reduction rule or a branching rule on $(G,k)$.
If $G$ does not contain an empty hyperedge, then $G'$ does not contain an empty hyperedge.
\end{lemma}
\begin{proof}
The lemma holds if $(G',k')$ is obtained from $(G,k)$ by applying Rule~(\ref{rule:loop}),
Rule~(\ref{rule:non-simple}), Rule~(\ref{rule:cc}), or Rule~(\ref{rule:not-connected}).
Now assume that $(G',k')$ is obtained by applying a different rule.
We have that $G' = \instance{G}{+x_1,\ldots,+x_s,-y_1,\ldots,-y_t}$ for some vertices
$x_1,\ldots,x_s,y_1,\ldots,y_t$.
Suppose for contradiction that $G'$ contains an empty hyperedge.
This implies that there is a hyperedge $e$ in $G$ such that $e \subseteq \{y_1,\ldots,y_t\}$.
Since Rule~(\ref{rule:loop}) cannot be applied on $(G,k)$, then $|e| \geq 2$,
and therefore $t \geq 2$.
Thus, the rule that was applied on $(G,k)$ is either (\ref{rule:deg2-2x}),
(\ref{rule:deg2-33-pre1}), (\ref{rule:deg2-33-pre2}), or (\ref{rule:deg2-33}).
If Rule~(\ref{rule:deg2-33-pre1}) was applied, then $t = 2$.
Since all the hyperedges in $G$ have size~3 when Rule~(\ref{rule:deg2-33-pre1}) is applied
(as Rules~(\ref{rule:loop}) and~(\ref{rule:d2}) cannot be applied on $(G,k)$),
we obtain that no hyperedge in $G$ is contained in $\{y_1,\ldots,y_t\}$,
a contradiction.
For the other rules we have that $\{y_1,\ldots,y_t\} = N(x)$ for some vertex $x$ with degree~2.
Since Rule~(\ref{rule:deg2-c}) cannot be applied on $(G,k)$, we have
that no hyperedge in $G$ is contained in $\{y_1,\ldots,y_t\}$,
a contradiction.
\end{proof}

We now give simple claims on the structure of instances on which some of the rules
above cannot be applied.
These claims are used to prove the correctness of the algorithm and to analyze the algorithm.

\begin{claim}\label{clm:deg-2-intersection}
If $(G,k)$ is an instance on which Rules~(\ref{rule:loop})--(\ref{rule:dominated}) cannot be applied
and $v$ is a vertex with degree~2, then the intersection of the two hyperedges that contain $v$
is $\{v\}$.
\end{claim}
\begin{proof}
Suppose for contradiction that the intersection of the two hyperedges that contain $v$
contains a vertex $w \neq v$.
The vertex $v$ is dominated by $w$, and therefore Rule~(\ref{rule:dominated}) can 
be applied, a contradiction.
\end{proof}

\begin{claim}\label{clm:min-deg-2}
If $(G,k)$ is an instance on which Rules~(\ref{rule:loop})--(\ref{rule:dominated}) cannot be applied,
then $G$ does not contain vertices with degree~1.
\end{claim}
\begin{proof}
Suppose for contradiction that $v$ is a vertex with degree~1.
Since Rule~(\ref{rule:loop}) cannot be applied, then
the unique hyperedge $e$ that contains $v$ has size at least~2.
Let $w$ be a vertex in $e \setminus \{v\}$.
The vertex $v$ is dominated by $w$, and therefore Rule~(\ref{rule:dominated}) can 
be applied, a contradiction.
\end{proof}

\begin{claim}\label{clm:dv}
If $(G,k)$ is an instance on which Rules~(\ref{rule:loop})--(\ref{rule:deg2-2x}) cannot be applied,
$m_2(G) > 0$, and $v$ is a vertex satisfying $d_2(v) = d_2(G)$, then
$d(v) \geq 3$.
\end{claim}
\begin{proof}
If $d_2(G) \geq 3$, then $d(v) \geq d_2(v) = d_2(G) \geq 3$.
Otherwise, $d_2(G) \in \{1,2\}$ (as $m_2(G) > 0$).
By Claim~\ref{clm:min-deg-2}, $d(v) \neq 1$.
We also have that $d(v) \neq 2$ since otherwise Rule~(\ref{rule:deg2-2x}) can be applied on $(G,k)$
and the vertex $v$.
Therefore, $d(v) \geq 3$.
\end{proof}

\begin{claim}\label{clm:m2-0}
If $(G,k)$ is an instance on which Rules~(\ref{rule:loop})--(\ref{rule:d2}) cannot be applied,
then all the hyperedges in $G$ have size~3.
\end{claim}
\begin{proof}
By Lemma~\ref{lem:empty}, $G$ does not contain empty hyperedges.
Since Rule~(\ref{rule:loop}) cannot be applied on $(G,k)$, $G$ does not contain 1-hyperedges.
Since Rule~(\ref{rule:d2}) cannot be applied on $(G,k)$, $G$ does not contain 2-hyperedges.
\end{proof}

\begin{claim}\label{clm:xy}
If $(G,k)$ is an instance on which Rules~(\ref{rule:loop})--(\ref{rule:deg2-33-pre1}) cannot be applied
and $d(G) \leq 4$, then for every two vertices $x,y$, there are at most two hyperedges
that contain both $x$ and $y$.
\end{claim}
\begin{proof}
If $d(G) = 4$, then the claim follows from the assumption that Rule~(\ref{rule:deg2-33-pre1}) cannot be
applied.
Suppose that $d(G) \leq 3$ and suppose for contradiction that there are vertices $x,y$
and hyperedges $e_1,e_2,e_3$ such that $x,y \in e_i$ for every $i$.
Since $d(G) \leq 3$, $e_1,e_2,e_3$ are all the hyperedges containing $x$.
Therefore, $x$ is dominated by $y$, a contradiction to the assumption that
Rule~(\ref{rule:dominated}) cannot be applied.
\end{proof}

\begin{claim}\label{clm:min-deg-3}
If $(G,k)$ is an instance on which Rules~(\ref{rule:loop})--(\ref{rule:deg2-33}) cannot be applied
and $d(G) \leq 4$, then every non-isolated vertex in $G$ has degree at least~3.
\end{claim}
\begin{proof}
Let $v$ be a non-isolated vertex.
By Claim~\ref{clm:min-deg-2}, $d(v) \geq 2$.
Since Rule~(\ref{rule:deg2-33}) cannot be applied, then $d(v) \geq 3$.
\end{proof}

In the next lemmas, we prove the safeness of Rules~(\ref{rule:cc}), (\ref{rule:deg2-u}),
(\ref{rule:cycle}), (\ref{rule:deg2-33-pre1}), and (\ref{rule:deg2-33-pre2}).
The safeness of the remaining rules is proved in~\cite[Lemma~63]{wahlstrom2007algorithms}.
We note that for some rules of our algorithm,
the choices of the vertices on which these rules operate are different than
the choices of the corresponding rules in~\cite{wahlstrom2007algorithms}.
However, this does not affect the safeness of these rules.

\begin{lemma}\label{lem:safe-cc}
Rule~(\ref{rule:cc}) is safe.
\end{lemma}
\begin{proof}
If $(G,k)$ is a yes-instance, then let $S$ be a minimum size hitting set of $G$.
Since $S \cap V(H)$ is a hitting set of $H$, then $|S \cap V(H)| \geq |S_H|$.
Therefore, $S \setminus V(H)$ is a hitting set of $G-H$
and $|S \setminus V(H)| = |S| - |S \cap V(H)| \leq k-|S_H|$.
Therefore, $(G-H,k-|S_H|)$ is a yes-instance.

For the opposite direction, suppose that $(G-H,k-|S_H|)$ is a yes-instance,
and let $S'$ be a minimum size hitting set of $G-H$.
The set $S' \cup S_H$ is a hitting set of $G$ of size at most $k$,
and therefore $(G,k)$ is a yes-instance.
\end{proof}

\begin{lemma}
Rule~(\ref{rule:deg2-u}) is safe.
\end{lemma}
\begin{proof}
To prove the safeness of the rule it suffices to show that there is a
minimum size hitting set of $G$ that contains $x$.
Let $S$ be a minimum size hitting set of $G$.
Assume that $x \notin S$ otherwise we are done.
There is no vertex $y$ such that $y \in e$ for every hyperedge $e$ that contain $x$,
since otherwise $x$ is dominated by $y$, a contradiction to the fact that Rule~(\ref{rule:dominated})
cannot be applied.
Therefore, $|S \cap N(x)| \geq 2$.
The set $S' = (S \setminus N(x)) \cup \{x,u\}$ is a hitting set of $G$
(every hyperedge $e$ that is hit by a vertex $p \in S \cap N(x)$ satisfies $e \cap N(x) \neq \emptyset$
and therefore either $x \in e$ or $u \in e$).
Since $|S'| \leq |S|$, we obtain that $S'$ is a minimum size hitting set of $G$ that contains $x$.
\end{proof}

\begin{lemma}\label{lem:safe-cycle}
Rule~(\ref{rule:cycle}) is safe.
\end{lemma}
\begin{proof}
To prove the safeness of the rule it suffices to show that there is a
minimum size hitting set $S$ of $G$ such that either
$\{v_1,v_3\} \subseteq S$ or $\{v_2,v_4\} \subseteq S$.
Let $S$ be a minimum size hitting set of $G$.
If $\{v_1,v_3\} \subseteq S$, then we are done.
If $\{v_1,v_3\} \not\subseteq S$, then assume without loss of generality that $v_1 \notin S$.
In order to hit the hyperedges $\{v_1,v_2\}$ and $\{v_1,v_4\}$,
$S$ must contain $v_2$ and $v_4$.
\end{proof}

\begin{lemma}\label{lem:deg2-33-pre1}
Rule~(\ref{rule:deg2-33-pre1}) is safe.
\end{lemma}
\begin{proof}
To prove the safeness of the rule it suffices to show that there is a
minimum size hitting set $S$ of $G$ such that $\{x,y\} \not\subseteq S$.
Let $S$ be a minimum size hitting set of $G$.
Suppose that $\{x,y\} \subseteq S$, otherwise we are done.
Since $x$ is not dominated by $y$, there is a hyperedge $e$ that contains $x$ and
does not contain $y$.
Let $v \in e \setminus \{x\}$.
Since $d(G) = 4$, then $e_1,e_2,e_3,e$ are all the hyperedges that contain $x$.
Therefore, $S' = (S \setminus \{x\}) \cup \{v\}$ is a minimum size hitting set of $G$
(since $y \in S'$ hits $e_1,e_2,e_3$ and $v$ hits $e$).
\end{proof}

\begin{lemma}\label{lem:deg2-33-pre2}
Rule~(\ref{rule:deg2-33-pre2}) is safe.
\end{lemma}
\begin{proof}
By Claim~\ref{clm:deg-2-intersection}, $e_1 \cap e_2 = \{x\}$,
and therefore $|N(x)| = 4$.
Since $I(x) = 0$, then for every hyperedge $e \in E(G) \setminus \{e_1,e_2\}$
such that $e \cap N(x) \neq \emptyset$ we have $|e \cap N(x)| = 1$,
and thus $e$ corresponds to an edge $\{e \cap N(x), e \setminus N(x)\}$ in the graph $H$.
Additionally, every element in $B$ is set of size~2.

For $p \in N(x)$, let $e_{p,1},e_{p,2}$ be the two hyperedges in $E(G) \setminus \{e_1,e_2\}$
that contain $p$.
Let $e'_{p,i} = e_{p,i} \setminus \{p\}$.
We have that $d^H(p) = 2$ for every $p \in N(x)$
(the neighbors of $p$ in $H$ are $e'_{p,1}$ and $e'_{p,2}$)
and therefore
$|E_H| = \sum_{p \in N(x)} d^H(p) = 2|N(x)| = 8$.
By Claim~\ref{clm:xy}, $d^H(e) \leq 2$ for every $e \in B$, and therefore
$|E_H| = \sum_{e \in B} d^H(e) \leq 2|B| = 8$.
Thus, $d^H(e) = 2$ for every $e \in B$.
This means that $H$ is a disjoint union of chordless cycles.
Since $|N(x)| = |B| = 4$, then either $H$ is the disjoint union of two chordless cycles of length 4,
or $H$ is a chordless cycle of length 8.

Assume that the first case of the rule occurs.
Namely, $H$ consists of a cycle $y,e'_{y,1},v,e'_{y,2},y$ and 
a cycle $z,e'_{z,1},w,e'_{z,2},z$.
To prove the safeness of the rule in this case it suffices to show that there is a
minimum size hitting set $S$ of $G$ such that either
(1) $x \in S$ and $y,z,v,w \notin S$,
(2) $y,v \in S$ and $x \notin S$, or
(3) $z,w \in S$ and $x \notin S$.
By Lemma~63 in~\cite{wahlstrom2007algorithms}, there is a minimum size hitting set $S$ of $G$
such that either $S$ satisfies (1) above, or $x \notin S$.
If $S$ satisfies (1), then we are done.
Suppose that $x \notin S$.
We claim that either $S$ satisfies (2) or $S$ satisfies (3).
Suppose for contradiction that $S$ does not satisfy (2) or (3).
Since $x \notin S$ and $S$ hits $e_1$, then $y \in S$ or $z \in S$.
Without loss of generality, $z \in S$.
By the assumption that $S$ does not satisfy (3) we have that $w \notin S$.
Since $x,w \notin S$ and $S$ hits $e_2$, then $v \in S$.
Therefore, $y \notin S$.
Since $S$ hits the hyperedges $e_{y,1},e_{y,2}$, then
$S$ contains a vertex $y_1 \in e'_{y,1}$ and a vertex $y_2 \in e'_{y,2}$.
Since $S$ hits the hyperedges $e_{w,1},e_{w,2}$, then
$S$ contains a vertex $w_1 \in e'_{w,1}$ and a vertex $w_2 \in e'_{w,2}$.
Note that the vertices $y_1,y_2,w_1,w_2$ are not necessarily distinct.
Recall that the hyperedges that contain $v$ are $e_2,e_{v,1},e_{v,2}$.
Since $y,e'_{y,1},v,e'_{y,2},y$ is a cycle in $H$, then
$\{e'_{y,1},e'_{y,2}\} = \{e'_{v,1},e'_{v,2}\}$.
Therefore, $\{y_1,y_2\}$ hits $e_{v,1}$ and $e_{v,2}$.
Since $z,e'_{z,1},w,e'_{z,2},z$ is a cycle in $H$, then
$\{w_1,w_2\}$ hits $e_{w,1}$ and $e_{w,2}$.
Therefore, $S' = (S \setminus \{z,v\}) \cup \{x\}$ is a hitting set of $G$.
Since $|S'| < |S|$, we obtain a contradiction.
Therefore, $S$ satisfies (2) or (3).

Assume that the first case of the rule occurs.
Namely, we have that either $H$ consists of a cycle $y,e'_{y,1},z,e'_{y,2},y$ and 
a cycle $v,e'_{v,1},w,e'_{v,2},v$,
or $H$ is a chordless cycle of length 8.
To prove the safeness of the rule in this case it suffices to show that there is a
minimum size hitting set $S$ of $G$ such that either
(1) $x \in S$ and $y,z,v,w \notin S$, or
(2) $N(x) \subseteq S$ and $x \notin S$.
By Lemma~63 in~\cite{wahlstrom2007algorithms}, there is a minimum size hitting set $S$ of $G$
such that either $S$ satisfies (1) above, or $x \notin S$.
If $S$ satisfies (1), then we are done.
Suppose that $x \notin S$.
Assume that $N(x) \not\subseteq S$, otherwise $S$ satisfies (2) and we are done.
Without loss of generality, $y \notin S$.
Therefore, $z \in S$.
Additionally, $S$ contains vertices $y_1 \in e'_{y,1}$ and $y_2 \in e'_{y,2}$.
If $v,w \in S$, then construct a set $S'$ from $S$ by removing the vertices $z,v,w$,
adding the vertex $x$, and adding one vertex from every set in $B \setminus \{e'_{y,1},e'_{y,2}\}$.
The set $S'$ is a hitting set of $G$.
Since $|B \setminus \{e'_{y,1},e'_{y,2}\}| = 2$, then $S'$ is a minimum hitting set of $G$.
The set $S'$ satisfies (1), so we are done.

Now suppose that $S$ contains exactly one vertex from $\{v,w\}$.
Without loss of generality, $v \notin S$ and $w \in S$.
The set $S$ contains vertices $v_1 \in e'_{v,1}$ and $v_2 \in e'_{v,2}$.
Let $E' = \{e'_{y,1},e'_{y,2},e'_{v,1},e'_{v,2}\}$.
Since the first case of the rule does not occur, $\{e'_{y,1},e'_{y,2}\} \neq \{e'_{v,1},e'_{v,2}\}$
and therefore $|E'| \geq 3$.
Construct a set $S'$ from $S$ by removing the vertices $z,w$, adding the vertex $x$,
and adding a vertex from the unique set in $B \setminus E'$ if $|E'| = 3$.
We again have that $S'$ is a minium size hitting set of $G$ and $S$ satisfies (1).
\end{proof}

\section{Analysis}

We first give several lemmas that will be used in the analysis of the algorithm.
A hypergraph $G$ is called \emph{good} if
(1) $G$ is a 3-hypergraph,
(2) $d(G) \leq 3$, and
(3) every connected component of $G$ contains a hyperedge of size at most 2
or a vertex with degree at most 2.

\begin{lemma}\label{lem:good-0}
Let $G$ be a connected 3-hypergraph such that $d(G) \leq 3$.
Let $G'$ be a hypergraph that is obtained from $G$ by deleting a vertex or a hyperedge of $G$.
Then, $G'$ is good.
\end{lemma}
\begin{proof}
It is clear that $G'$ satisfies (1) and (2).
Let $H_1,\ldots,H_s$ be the connected components of $G'$.
We need to show that each component $H_i$ contains a hyperedge of size at most 2
or a vertex with degree at most 2.

If $G'$ is obtained from $G$ by deleting a hyperedge $e$, then every component $H_i$
contains a vertex $x_i$ such that $x_i \in e$.
We have that $d^{G'}(x_i) = d^{G}(x_i)-1 \leq 2$.
If $G'$ is obtained from $G$ by deleting a vertex $x$, then every component $H_i$
contains a hyperedge $e_i$ such that $e_i \cup \{x\}$ is a hyperedge of $G$.
We have that $|e_i| = |e_i \cup \{x\}|-1 \leq 2$.
Therefore, $G'$ satisfies (3).
\end{proof}

\begin{lemma}\label{lem:good}
Let $G$ be a good hypergraph.
Let $G'$ be a hypergraph that is obtained from $G$ by deleting a vertex or a hyperedge of $G$.
Then, $G'$ is good.
\end{lemma}
\begin{proof}
It is clear that $G'$ satisfies (1) and (2).
Let $H$ be the connected component of $G$ that contains the deleted hyperedge or deleted vertex.
Every connected component $H' \neq H$ of $G$ is also a connected component of $G'$,
and since $G$ is good,
$H'$ contains a hyperedge of size at most 2 or a vertex of degree at most 2.
Let $H_1,\ldots,H_s$ be the connected components of $G'$ that contain vertices of $H$.
By applying Lemma~\ref{lem:good-0} on $H$ we obtain that
every component $H_i$ contains a hyperedge of size at most 2 or a vertex with degree at most 2.
Therefore, $G'$ satisfies (3).
\end{proof}

\begin{lemma}\label{lem:path}
Let $G$ be a 3-hypergraph such that $d(G) \leq 3$, and let $k$ be a nonnegative integer.
Then, in every path from the root to a leaf in the recursion tree of $(G,k)$ there is
at most one node that corresponds to an application of Rule~(\ref{rule:final}).
\end{lemma}
\begin{proof}
Let $x_1,\ldots,x_s$ be a path from the root to a leaf in the recursion tree of $(G,k)$.
Let $(G_i,k_i)$ be the instance corresponding to $x_i$.
Let $i$ be the minimal index such that Rule~(\ref{rule:final}) is applied on
$(G_i,k_i)$ (if $i$ does not exist then we are done).
Since $d(G_i) \leq d(G) \leq 3$ and Rule~(\ref{rule:not-connected}) cannot be applied on $G_i$,
we have that $G_i$ is connected.
The hypergraph $G_{i+1}$ is obtained from $G_i$ by deleting one vertex and some hyperedges.
By Lemma~\ref{lem:good-0} and Lemma~\ref{lem:good}, the hypergraph $G_{i+1}$ is good.
By Lemma~\ref{lem:good}, $G_j$ is good for every $j \geq i+1$
(as $G_j$ is obtained from $G_{j-1}$ by deleting vertices and hyperedges).
Consider some instance $(G_j,k_j)$ for some $j \geq i+1$.
Since $G_j$ is good, either $G_j$ contain a hyperedge of size at most 2
or a vertex with degree at most 2.
In the former case, by Claim~\ref{clm:m2-0},
the algorithm applies Rule~(\ref{rule:d2}) on $(G_j,k_j)$ or an earlier rule.
In the latter case, by Claim~\ref{clm:min-deg-3},
the algorithm applies Rule~(\ref{rule:deg2-33}) on $(G_j,k_j)$ or an earlier rule.
Therefore, Rule~(\ref{rule:final}) is not applied on $(G_j,k_j)$ for every $j \geq i+1$.
\end{proof}

To analyze the time complexity of algorithm \hsb, we use the measure and conquer approach.
We use functions $\Psi_3,\Psi_4,\Psi_5,\Psi_6$ from $\mathbb{N} \times \mathbb{N}$
to $\mathbb{R}$ and we assume that these functions have the following properties:

\begin{enumerate}[(P1)]
\item\label{prop-range}
$0 \leq \Psi_i(m,c) \leq 2$ for every $m \geq 0$, $c \geq 0$, and $i \in [3,5]$.
Additionally,
$0 \leq \Psi_6(m,c) \leq 1$ for every $m \geq 0$ and $c \geq 0$.

\item\label{prop-0}
$\Psi_i(0,0) = 0$ for every $i \in [3,6]$.

\item\label{prop-monotone}
$\Psi_i(m+1,c') \geq \Psi_i(m,c)$ for every $m \geq 1$, $c \in [1,m]$, $c' \in [1,m+1]$,
and $i \in [3,6]$.

\item\label{prop-bounded}
$\Psi_i(m,c)-\Psi_i(m-a,c') \leq 1$ for every $m \geq 1$, $a \leq \min(i,m)$, $c \in [1,m]$, $c' \in [1,m-a]_0$,
and $i \in [3,5]$.
\end{enumerate}
Let $\mu_i(G,k) = k - \Psi_i(m_2(G),c_2(G))$.
The measure function that is used for the analysis of the algorithm is
$\mu(G,k) = \mu_{\hat{d}(G)}(G,k)$,
where $\hat{d}(G) = d(G)$ if $3 \leq d(G) \leq 6$,
$\hat{d}(G) = 3$ if $d(G) < 3$, and
$\hat{d}(G) = 6$ if $d(G) > 6$.
We note that this measure function is similar to the one used in~\cite{wahlstrom2007algorithms}
(the difference between the functions is that the measure function in~\cite{wahlstrom2007algorithms}
does not depend on $c_2(G)$).

When considering the branching vectors of some branching rule of the algorithm,
we do not take into account that the hypergraphs generated by the rule can have
different $\hat{d}(\cdot)$ values than $G$.
That is, if the rule generates instances $(G_1,k_1),\ldots,(G_s,k_s)$
from an instance $(G,k)$,
then we assume that the branching vector is
$(\mu_d(G,k)-\mu_d(G_1,k_1),\ldots,\mu_d(G,k)-\mu_d(G_s,k_s))$,
where $d = \hat{d}(G)$.
In Lemma~\ref{lem:leaves} we will take into account the changes in $\hat{d}(\cdot)$.
Informally, the reason that the assumption above is justified is that along
a path in the recursion tree, the value of $\hat{d}(\cdot)$ changes at most three times
(since the value $\hat{d}(\cdot)$ does not increase along the path),
and each change in the value of $\hat{d}(\cdot)$ yields a bounded
change in the measure function due to Property~(P\ref{prop-range}).

Property~(P\ref{prop-monotone}) is used in the analysis in the following way.
Suppose that we consider a branching rule that generates instances
$(G_1,k_1),\ldots,(G_s,k_s)$ from an instance $(G,k)$.
The analysis of the rule requires giving a lower bound on
$\mu_d(G,k)-\mu_d(G_i,k_i)$ for every $i \in [1,s]$.
If we can show that $m_2(G_i) \geq m'$ for some integer $m'$,
then we can use the following claim to give a lower bound
on $\mu_d(G,k)-\mu_d(G_i,k_i)$.
\begin{claim}\label{clm:mu}
Let $(G,k)$ and $(G',k')$ be two instances.
If $m_2(G') \geq m'$ for some integer $m'$, then
$\mu_d(G,k)-\mu_d(G',k') \geq (k-k')-(\Psi_d(m_2(G),c_2(G)) - \Psi_d(m',c'))$,
where $d = \hat{d}(G)$ and $c' = \min(m',c_2(G'))$.
\end{claim}
\begin{proof}
By definition,
$\mu_d(G,k)-\mu_d(G',k') = (k-k')-(\Psi_d(m_2(G),c_2(G)) - \Psi_d(m_2(G'),c_2(G')) )$.
By Property~(P\ref{prop-monotone}) we have
$\Psi_d(m_2(G_i),c_2(G_i)) \geq \Psi(m',c')$,
and the claim follows.
\end{proof}

Property~(P\ref{prop-bounded}) is used to prove the following lemma.
\begin{lemma}\label{lem:reduction}
Let $(G,k)$ be an instance on which algorithm \hsb\ applies Rule (R$i$) for some $i \geq 1$.
Let $(G',k')$ be the instance obtained by applying Rule~(R$i$) on $(G,k)$
and then exhaustively applying Rule~(\ref{rule:loop}) if Rule~(R$i$) created new 1-hyperedges.
Then, $\mu_d(G',k') \leq \mu_d(G,k)$, where $d = \hat{d}(G)$.
\end{lemma}
\begin{proof}
Let $\Delta_k = k-k'$,
$\Delta_\Psi = \Psi_d(m_2(G),c_2(G)) - \Psi_d(m_2(G'),c_2(G'))$,
and $\Delta_m = m_2(G)-m_2(G')$.
By definition, $\mu_d(G',k') \leq \mu_d(G,k)$ if and only if $\Delta_\Psi \leq \Delta_k$.
We claim that one of the following cases occurs.
\begin{enumerate}
\item
$\Delta_k = 0$, $\Delta_m = 0$, and $c_2(G) = c_2(G')$.
\item
$\Delta_k = 0$ and $\Delta_m < 0$.
\item
$\Delta_k \geq 1$ and $\Delta_m \leq d(G) \cdot \Delta_k$.
\end{enumerate}
If the algorithm applies Rule~(\ref{rule:cc}) on $(G,k)$,
then case~3 occurs since in this case $\Delta_k = |S_H| \geq 1$,
every 2-hyperedge in $E(G) \setminus E(G')$ contains a vertex from $S_H$,
and a vertex from $S_H$ is contained in at most $d(G)$ 2-hyperedges.
If the algorithm applies Rule~(\ref{rule:deg2-u}) on $(G,k)$,
then case~3 occurs since in this case $\Delta_k = 1$ and $\Delta_m \leq d(G)$.
The claim holds for the other reduction rules by the proof of Lemma~64 in~\cite{wahlstrom2007algorithms}.

If case~1 above occurs, then $\Delta_\Psi = 0 = \Delta_k$.
If case~2 occurs, then by Property~(P\ref{prop-monotone}), $\Delta_\Psi \leq 0 = \Delta_k$.
Suppose that case~3 occurs.
If $\Delta_k = 1$, then by Property~(P\ref{prop-range}) (if $d = 6$)
and Property~(P\ref{prop-bounded}) (if $d < 6$) we have $\Delta_\Psi \leq 1 = \Delta_k$.
Otherwise, by Property~(P\ref{prop-range}), $\Delta_\Psi \leq 2 \leq \Delta_k$.
\end{proof}

In the following, we write $\Psi$ instead of $\Psi_{\hat{d}(G)}$.
We also define
$\Delta(m,\alpha,c,c') = \Psi(m,c)-\Psi(\max(0,m-\alpha),c')$
and
$\Psi(m,*) = \min_{c \in [1,m]_0} \Psi(m,c)$.

Let $G,G'$ be two hypergraphs on the same set of vertices.
Define
$\rem{G}{G'} = E_2(G) \setminus E_2(G')$ and
$\add{G}{G'} = E_2(G') \setminus E_2(G)$.
Clearly, $m_2(G') = m_2(G)-|\rem{G}{G'}|+|\add{G}{G'}|$.
The following observation will be used in the analysis of the algorithm.

\begin{observation}\label{clm:m2}
Let $G$ be a 3-hypergraph and let $G' = \instance{G}{+x_1,\ldots,+x_s,-y_1,\ldots,-y_t}$
for some vertices $x_1,\ldots,x_s,y_1,\ldots,y_t$.
Then,
\[ \rem{G}{G'} = \{e : e \in E_2(G), e \cap\{x_1,\ldots,x_s,y_1,\ldots,y_t\} \neq \emptyset \} \]
and
\[ \add{G}{G'} = \{e \setminus \{y_1,\ldots,y_t\} :
  e \in E_3(G), e \cap \{x_1,\ldots,x_s\} = \emptyset, |e \cap \{y_1,\ldots,y_t\}| = 1\}. \]
\end{observation}

\subsection{Analysis of Rules~(\ref{rule:cycle})--(\ref{rule:deg2-33})}
\label{sec:analysis-branching-rules}

We note that for some rules we can give better bounds on their branching vectors.
However, since our simple bounds already show that these rules have branching
numbers smaller than 2.04, we only give the simple bounds.

\begin{lemma}\label{lem:rule-cycle}
The branching vectors of Rule~(\ref{rule:cycle}) are dominated by
\[ \{(2-\Delta(m,4,c,c-1), 2-\Delta(m,4,c,c-1)) : m \geq 4, c \in [2,m-3]_1\}. \]
\end{lemma}
\begin{proof}
By the definition of the rule, $m_2(G) \geq 4$.
Since $d_2(G) = 2$, then $G^2$ contains a connected component $H$ such that $V(H) = \{v_1,\ldots,v_4\}$
and $H$ is a chordless cycle of length 4.
Therefore, $c_2(G) \in [2,m_2(G)-3]_1$.

Let $G_1 = \instance{G}{+v_1,+v_3}$ and $G_2 = \instance{G}{+v_2,+v_4}$.
Fix $i \in \{1,2\}$.
By Observation~\ref{clm:m2},
$\rem{G}{G_i} = \{\{v_1,v_2\}, \{v_2,v_3\}, \{v_3,v_4\}, \{v_1,v_4\}\}$ and
$\add{G}{G_i} = \emptyset$.
Therefore, $m_2(G_i) = m_2(G)-4$.
The graph $(G_i)^2$ is obtained from $G^2$ by deleting the connected component $H$.
Thus, $c_2(G_i) = c_2(G)-1$.
It follows that the branching vector for the instance $(G,k)$ is
$(2-\Delta(m,4,c,c-1), 2-\Delta(m,4,c,c-1))$, where
$m = m_2(G)$ and $c = c_2(G)$.
\end{proof}

\begin{figure}
\centering
\subfigure[\label{fig:deg2-2-1}]{\includegraphics{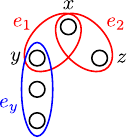}}%
\qquad%
\subfigure[\label{fig:deg2-2-2}]{\includegraphics{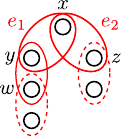}}%
\qquad%
\subfigure[\label{fig:deg2-2-3}]{\includegraphics{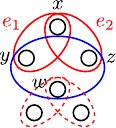}}%
\caption{The cases of Lemma~\ref{lem:rule-deg2-2x-2}.
Ellipses with solid lines are hyperedges that must exist,
and ellipses with dashed lines are 2-hyperedges that may exist.}
\end{figure}

\begin{lemma}\label{lem:rule-deg2-2x-2}
The branching vectors of the first case of Rule~(\ref{rule:deg2-2x}) are dominated by
\begin{align*}
& \{(1-\Delta(m,1,c,c'),2-\Delta(m,2,c,c-1)) : m \geq 2, c \in [2,m-1]_1, c' \in [1,m-1]\} \cup\\
& \{(2-\Delta(m,5,c,c'),2-\Delta(m,4,c,c'')) : m \geq 2, c \in [1,m-1], c' \in [1,m-5]_0, c'' \in [1,m-4]_0\}.
\end{align*}
\end{lemma}
\begin{proof}
Since $e_1$ and $e_2$ are 2-hyperedges, we have $m_2(G) \geq 2$.
Additionally, $e_1$ and $e_2$ are in the same connected component of $G^2$, and therefore
$c_2(G) \leq m_2(G)-1$.
Let $G_1 = \instance{G}{+x,-y,-z}$ and $G_2 = \instance{G}{-x,+y,+z}$.

We first assume that $d_2(y) = d_2(z) = 1$ and $I(x) = 0$
(see Figure~\ref{fig:deg2-2-1}).
Due to the assumption $d_2(y) = d_2(z) = 1$ we have that $\rem{G}{G_1} = \{e_1,e_2\}$.
Since $y$ is not dominated by $x$, there is a hyperedge $e$ that contains $y$
and does not contain $x$.
Since $e_1$ is the only 2-hyperedge that contains $y$, then $|e| = 3$.
We have that $z \notin e$ due to the assumption that $I(x) = 0$.
Therefore, $e \setminus \{y\} \in \add{G}{G_1}$.
We showed that $|\rem{G}{G_1}| = 2$ and $|\add{G}{G_1}| \geq 1$ and therefore
$m_2(G_1) \geq m_2(G)-1$.
Due to the assumption that $d_2(y) = d_2(z) = 1$ we have that $\rem{G}{G_2} = \{e_1,e_2\}$.
Additionally, $\add{G}{G_2} = \emptyset$.
Therefore, $m_2(G_2) = m_2(G)-2$.
Since $d_2(y) = d_2(z) = 1$, then $c_2(G_2) = c_2(G)-1$
($G^2$ contains a connected component $H$ whose vertices are $x,y,z$,
and $(G_2)^2$ is obtained from $G$ by deleting $H$).
By claim~\ref{clm:mu}, the branching vector for the instance $(G,k)$ is dominated by
$(1-\Delta(m,1,c,c'), 2-\Delta(m,3,c,c-1))$, where
$m = m_2(G)$, $c = c_2(G)$, and $c' = \min(m-1,c_2(G_1))$.
Note that $m \geq 2$, $c \in [2,m-1]_1$, and $c' \in [1,m-1]$.

Now suppose that the case above does not occur, so either $d_2(y) > 1$, $d_2(z) > 1$,
or $I(x) = 1$.
Since $d_2(G) = 2$, then either
(1) $d_2(y) = 2$
(2) $d_2(z) = 2$ or
(3) $d_2(y) = d_2(z) = 1$ and $I(x) = 1$.
Without loss of generality we assume that either (1) or (3) occurs.
Namely, either there is a hyperedge $\{y,w\}$ (see Figure~\ref{fig:deg2-2-2})
or $d_2(y) = d_2(z) = 1$ and there is a hyperedge $e$ that contains $y$ and $z$
(see Figure~\ref{fig:deg2-2-3}).
Since  Rule~(\ref{rule:deg2-c}) cannot be applied on $(G,k)$, then
in the former case $w \neq z$ and in the latter case $|e| = 3$.
In the latter case we denote $e = \{y,z,w\}$.
In both cases, the hypergraph $G_1$ contains a 1-hyperedge $\{w\}$.
Therefore, Rule~(\ref{rule:loop}) is applied on the instance $(G_1,k-1)$.
The resulting instance is $(G'_1,k-2)$ where $G'_1 = \instance{G}{+x,-y,-z,+w}$.
We assume that the application of Rule~(\ref{rule:loop}) occurs inside Rule~(\ref{rule:deg2-2x}).
In other words, we assume that Rule~(\ref{rule:deg2-2x}) generates the instances
$(G'_1,k-2)$ and $(G_2,k-2)$.
Since $d_2(G) = 2$, we have that $d_2(y) \leq 2$, $d_2(z) \leq 2$, and $d_2(w) \leq 2$.
It follows that $|\rem{G}{G'_1}| \leq 5$
(If case~(1) occurs, then
$\rem{G}{G'_1}$ consists of $e_1$, $e_2$, $\{y,w\}$,
at most one 2-hyperedge that contains $z$ and not $x$, and
at most one 2-hyperedge that contains $w$ and not $y$.
If case~(3) occurs, then $\rem{G}{G'_1}$ consists of $e_1$, $e_2$,
and at most two 2-hyperedges that contain $w$).
Therefore, $m_2(G'_1) \geq m_2(G) - 5$.
Additionally, $\rem{G}{G_2}$ consists of at most two 2-hyperedges that contain $y$ and
at most two 2-hyperedges that contain $z$, so $m_2(G_2) \geq m_2(G) - 4$.
By claim~\ref{clm:mu}, the branching vector for the instance $(G,k)$ is dominated by
$(2-\Delta(m,5,c,c'),2-\Delta(m,4,c,c''))$
for some $m \geq 2$, $c \in [1,m]$, $c' \in [1,m-5]_0$, and $c'' \in [1,m-4]_0$.
\end{proof}

\begin{figure}
\centering
\subfigure[\label{fig:deg2-2-4}]{\includegraphics{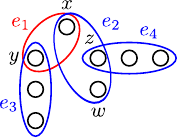}}%
\qquad%
\subfigure[\label{fig:deg2-2-5}]{\includegraphics{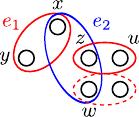}}%
\qquad%
\subfigure[\label{fig:deg2-2-6}]{\includegraphics{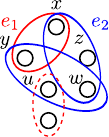}}%
\caption{The cases of Lemma~\ref{lem:rule-deg2-2x-1}.}
\end{figure}

\begin{lemma}\label{lem:rule-deg2-2x-1}
The branching vectors of the second case of Rule~(\ref{rule:deg2-2x}) are dominated by
\begin{align*}
& \{(1-\Delta(m,-1,m,c'),1) : m \geq 1, c' \in[1,m+1]\} \cup \\
& \{(2-\Delta(m,3,m,c'),1-\Delta(m,0,m,c'')) : m \geq 1, c' \in [1,m-3]_0, c' \in [1,m]\}.
\end{align*}
\end{lemma}
\begin{proof}
By definition, every connected component of $G^2$ consists of one edge,
and therefore $c_2(G) = m_2(G)$.
Let $G_1 = \instance{G}{+x,-y,-z,-w}$ and $G_2 = \instance{G}{-x,+y}$.
Since $d_2(G) = 1$, then $d_2(y) = 1$.
Therefore, $\rem{G}{G_2} = \{e_1\}$. 
Additionally, $\add{G}{G_2} = \{e_2 \setminus \{x\}\}$.
It follows that $m_2(G_2) = m_2(G)$.

We first assume that $d_2(z) = d_2(w) = 0$ and $I(x) = 0$ (see Figure~\ref{fig:deg2-2-4}).
We have $\rem{G}{G_1} = \{e_1\}$ (since $d_2(y) = 1$ and $d_2(z) = d_2(w) = 0$).
Let $e_3$ be a hyperedge such that $x \notin e_3$ and $e_3 \cap N(x) = \emptyset$
($e_3$ exists since $d(y) \geq 2$ by Claim~\ref{clm:min-deg-2}).
Since $I(x) = 0$ and Rule~(\ref{rule:deg2-c}) cannot be applied on $(G,k)$, then
$|e_3 \cap N(x)| = 1$, and therefore $e_3 \setminus N(x) \in \add{G}{G_1}$.
Let $u \in e_3 \setminus N(x)$.
Since Rule~(\ref{rule:deg2-u}) cannot be applied,
there is a hyperedges $e_4$ such that 
$x,u \notin e_4$ and $e_4 \cap N(x) \neq \emptyset$.
Since $I(x) = 0$ and Rule~(\ref{rule:deg2-c}) cannot be applied on $(G,k)$, then
$e_4 \setminus N(x) \in \add{G}{G_1}$.
Since $u \notin e_4$, then $e_3 \setminus N(x) \neq e_4 \setminus N(x)$.
Therefore, $|\add{G}{G_1}| \geq 2$ and $m_2(G_1) \geq m_2(G) + 1$.

Recall that every connected component in $G^2$ consists of one edge.
Since $d_2(z) = d_2(w) = 0$, then
the graph $(G_2)^2$ is obtained from $G^2$ by removing the vertices $x,y$ and the edge $\{x,y\}$,
and adding new vertices $z,w$ (since $d_2(z) = d_2(w) = 0$) and an edge $\{z,w\}$.
Therefore, $c_2(G_2) = c_2(G)$, so
$\Psi(m_2(G_2),c_2(G_2)) = \Psi(m_2(G),c_2(G))$
(recall that $m_2(G_2) = m_2(G)$).
By claim~\ref{clm:mu}, the branching vector for the instance $(G,k)$ is dominated by 
$(1-\Delta(m,-1,m,c'),1)$ for some $m \geq 1$ and $c' \in [1,m+1]$.

Now suppose that $d_2(z) \geq 1$ or $d_2(w) \geq 1$ (see Figure~\ref{fig:deg2-2-5}).
Without loss of generality, $d_2(z) \geq 1$.
Since $d_2(G) = 1$, then $d_2(z) = 1$.
Let $\{z,u\}$ be the unique 2-hyperedge that contains $z$.
Note that $u \neq x$ (since $e_1$ and $e_2$ are the only hyperedges that contain $x$),
$u \neq y$ (since Rule~(\ref{rule:deg2-c}) cannot be applied on $(G,k)$), and
$u \neq w$ (since Rule~(\ref{rule:non-simple}) cannot be applied on $(G,k)$).
The hypergraph $G_1$ contains a 1-hyperedge $\{u\}$.
Therefore, Rule~(\ref{rule:loop}) is applied on the instance $(G_1,k-1)$.
As before, we assume that Rule~(\ref{rule:deg2-2x}) generates the instances $(G'_1,k-2)$ and $(G_2,k-1)$,
where $G'_1 = \instance{G}{+x,-y,-z,-w,+u}$.
The set $\rem{G}{G'_1}$ consists of $e_1$, $\{z,u\}$, and at most one 2-hyperedge that contains $w$
(since $d_2(y) = d_2(z) = d_2(u) = 1$ and $d_2(w) \leq 1$).
Thus, $m_2(G'_1) \geq m_2(G) - 3$.
the graph $(G_1)^2$ is obtained from $G^2$ by removing the
edges in $\rem{G}{G'_1}$ and removing their vertices.
Since every connected component in $G^2$ is an edge, we obtain that
$c_2(G'_1) \geq c_2(G) - |\rem{G}{G'_1}| = m_2(G) - |\rem{G}{G'_1}|$.
Recall that $m_2(G_2) = m_2(G)$.
By claim~\ref{clm:mu}, the branching vector for the instance $(G,k)$ is dominated by 
$(2-\Delta(m,3,m,c'),1-\Delta(m,0,m,c''))$
for some $m \geq 2$, $c' \in [1,m-3]_0$, and $c'' \in [1,m]$.

Finally, suppose that $d_2(z) = d_2(w) = 0$ and $I(x) = 1$.
Namely, there is a hyperedge $e$ such that $|e \cap N(x)| \geq 2$ (see Figure~\ref{fig:deg2-2-6}).
Since Rule~(\ref{rule:deg2-c}) cannot be applied on $(G,k)$, then $|e \cap N(x)| = 2$.
Let $u$ be the unique vertex in $e \setminus N(x)$.
Note that $u \neq x$ (since $e_1$ and $e_2$ are the only hyperedges that contain $x$).
Since $\{u\}$ is a 1-hyperedge in $G_1$, we assume that Rule~(\ref{rule:deg2-2x}) generates
the instances $(G'_1,k-2)$ and $(G_2,k-1)$, where $G'_1 = \instance{G}{+x,-y,-z,-w,+u}$.
The set $\rem{G}{G'_1}$ consists of $e_1$ and at most one 2-hyperedge that contains $u$
(since $d_2(y) = 1$, $d_2(z) = d_2(w) = 0$, and $d_2(u) \leq 1$).
Thus, $m_2(G'_1) \geq m_2(G) - 2 > m_2(G)-3$.
By claim~\ref{clm:mu}, the branching vector for the instance $(G,k)$ is dominated by 
$(2-\Delta(m,3,m,c'),1-\Delta(m,0,m,c''))$
for some $m \geq 1$, $c' \in [1,m-3]_0$, and $c'' \in [1,m]$.
\end{proof}

\begin{lemma}\label{lem:rule-d2}
The branching vectors of Rule~(\ref{rule:d2}) on instances $(G,k)$ with $d_2(G) = 1$
are dominated by
\[ \{(1-\Delta(m,1,m,m-1),1-\Delta(m,-1,m,c')) :  m \geq 1, c' \in [1,m+1]\}. \]
The branching vectors of Rule~(\ref{rule:d2}) on instances $(G,k)$ with $d_2(G) = 1$
in which $d_3(x) \geq 3$ are dominated by
\[ \{(1-\Delta(m,1,m,m-1),1-\Delta(m,-2,m,c')) :  m \geq 1, c' \in [1,m+2]\}. \]
The branching vectors of Rule~(\ref{rule:d2}) on instances $(G,k)$ with $d_2(G) = 2$
are dominated by
\begin{align*}
& \{(1-\Delta(m,2,c,c'),2-\Delta(m,\min(m,4)-1,c,c'')) : \\
&   \qquad m \geq 2, m \neq 4, c \in [1,m-1], c' \in [1,m-2]_0, c'' \in [1,m-3]_1\} \cup\\
& \{(1-\Delta(4,2,c,c'),2-\Delta(4,2,c,c'')) : c \in [1,3],c' \in [1,2],c'' \in [1,2]\}.
\end{align*}
For every $d \geq 3$,
the branching vectors of Rule~(\ref{rule:d2}) on instances $(G,k)$ with $d_2(G) = d$
are dominated by
\begin{align*}
& \{(1-\Delta(m,d,c,c'),d-\Delta(m,\min(m,d^2),c,c'')) : \\
&   \qquad m \geq d, c \in [1,m-d+1], c' \in [1,m-d]_0, c'' \in [1,m-\min(m,d^2)]_0\}.
\end{align*}
\end{lemma}
\begin{proof}
Let $d = d_2(G)$ and
$e_1 = \{x,y_1\},\ldots,e_d = \{x,y_d\}$ be the 2-hyperedges that contain $x$.
The hypergraph $\instance{G}{-x}$ contains 1-hyperedges $\{y_i\}$ for $i \in [1,d]$.
Therefore, the algorithm applies Rule~(\ref{rule:loop}) $d$ times starting from
the instance $(\instance{G}{-x},k)$.
The resulting instance is $(G_2 = \instance{G}{-x,+y_1,\ldots,+y_d},k-d)$.
We assume that all these applications of Rule~(\ref{rule:loop}) occur inside Rule~(\ref{rule:d2}).
In other words, we assume that Rule~(\ref{rule:d2}) generates the instances
$(G_1 = \instance{G}{+x},k-1)$ and $(G_2,k-d)$.

We have that $\rem{G}{G_1} = \{e_1,\ldots,e_d\}$ and
$\add{G}{G_1} = \emptyset$.
Therefore, $m_2(G_1) = m_2(G)-d$.

Every hyperedge in $\rem{G}{G_2}$ contains a vertex from $y_1,\ldots,y_d$.
Therefore, $|\rem{G}{G_2}| \leq \min(m_2(G),\sum_{i=1}^d d_2(y_i)) \leq \min(m_2(G), d^2)$
(as $d_2(y_i) \leq d$ for all $i$).
Additionally, $\add{G}{G_2}$ consists of the $d_3(x)$ 3-hyperedges that contain $x$
(note that since $G$ is simple, for every 3-hyperedge $e$ that contains $x$
we have $y_i \notin e$ for all $i$).
Therefore, $m_2(G_2) \geq m_2(G) - \min(m_2(G), d^2) + d_3(x)$.
By Claim~\ref{clm:dv}, $d_3(x) = d(x)-d_2(x) \geq 3-d$, so
$m_2(G_2) \geq m_2(G) - \min(d^2, m_2(G)) + \max(0,3-d)$.

Consider the case $d = 1$.
From the previous paragraph we have $m_2(G_2) \geq m_2(G)-1+2 = m_2(G)+1$.
Every connected component of $G^2$ is an edge, and therefore $c_2(G) = m_2(G)$.
Additionally, the graph $(G_1)^2$ is obtained from $G^2$ by deleting the vertices $x,y_1$
and the edge $\{x,y_1\}$.
Therefore, $c_2(G_1) = m_2(G) - 1$.
By claim~\ref{clm:mu}, the branching vector for the instance $(G,k)$ is dominated by
$(1-\Delta(m,1,m,m-1),1-\Delta(m,-1,m,c'))$
for some $m \geq 1$ and $c' \in [1,m+1]$.
Moreover, if $d_3(x) \geq 3$, then $m_2(G_2) \geq m_2(G)-1+3 = m_2(G)+2$.
and the branching vector for the instance $(G,k)$ is dominated by
$(1-\Delta(m,1,m,m-1),1-\Delta(m,-2,m,c'))$ for some $m \geq 1$ and $c' \in [1,m+2]$.

We now assume that $d > 1$.
In this case, $e_1,\ldots,e_d$ are in the same connected component of $G^2$
and thus $c_2(G) \leq m-d+1$.
We have that $d_3(x) \geq 1$ if $d = 2$.
By claim~\ref{clm:mu}, if $d = 2$ then the branching vector for the instance $(G,k)$ is dominated by 
$(1-\Delta(m,2,c,c'),2-\Delta(m,\min(m,4)-1,c,c''))$
for some $m \geq 2$, $c \in [1,m-1]$, $c' \in [1,m-2]_0$, and $c'' \in [1,m-3]_1$.
If $d \geq 3$, then the branching vector is dominated by
$\{(1-\Delta(m,d,c,c'),d-\Delta(m,\min(m,d^2),c,c''))$
for some $m \geq d$, $c \in [1,m-d+1]$, $c' \in [1,m-d]_0$, and $c'' \in [1,m-\min(m,d^2)]_0$.

For the case $d = 2$ and $m_2(G) = 4$ we can give a better bound on the branching vector.
We will show that $m_2(G_2) \geq 2$.
If $G^2$ is not connected, then every connected component of $G^2$ contains at most $3$ edges.
Since the set $\rem{G}{G_2}$ contains 2-hyperedges of $G$ that are in the same connected component of $G^2$,
it follows that $|\rem{G}{G_2}| \leq 3$, and therefore $m_2(G) \geq 4-3+1=2$.
Now suppose that $G^2$ is connected.
Since $d_2(G) = 2$, the maximum degree of $G^2$ is 2.
Since Rule~(\ref{rule:cycle}) cannot be applied, $G^2$ is not a chordless cycle on 4 vertices.
Therefore, $G^2$ is a path on 5 vertices.
Denote this path $v_1,\ldots,v_5$.
By Claim~\ref{clm:dv}, $d(v_2) \geq 3$ and therefore $d_3(v_2) = d(v_2)-2 \geq 1$.
Additionally, we have $D_2(v_2) = 3$.
Thus, $d_3(v_2)-D_2(v_2) \geq 1-3 = -2$.
By the definition of $x$ in Rule~(\ref{rule:d2}), 
$d_3(x)-D_2(x) \geq d_3(v_2)-D_2(v_2) \geq -2$.
Since $|\rem{G}{G_2}| = D_2(x)$ and $|\add{G}{G_2}| = d_3(x)$,
it follows that $m_2(G) \geq 4-2 = 2$.
By claim~\ref{clm:mu}, the branching vector for the instance $(G,k)$ is dominated by
$(1-\Delta(4,2,c,c'),2-\Delta(4,2,c,c''))$
for some $c \in [1,3]$, $c' \in [1,2]$, and $c'' \in [1,2]$.
\end{proof}

In the following lemmas we analyze Rules~(\ref{rule:deg2-33-pre1}), (\ref{rule:deg2-33-pre2}),
and Rule~(\ref{rule:deg2-33}).
When these rules are applied we have that $m_2(G) = 0$ by Claim~\ref{clm:m2-0}.
Therefore, the elements in the branching vectors of these rules have the
form $a+\Psi(m,c)$:
If $(G',k')$ is one of the instances generated from $(G,k)$ by one of these rules,
then the corresponding value in the branching vector is
$\mu_d(G,k)-\mu_d(G',k') = (k-k') + \Psi(m_2(G'),c_2(G'))-\Psi(m_2(G),c_2(G))
= (k-k') + \Psi(m_2(G'),c_2(G'))$, where the last equality follows from Property~(P\ref{prop-0}).
In some cases we use the fact that $\Psi(m_2(G'),c_2(G')) \geq 0$
(Property~(P\ref{prop-range})) and the bound becomes
$\mu_d(G,k)-\mu_d(G',k') \geq k-k'$.

\begin{lemma}\label{lem:rule-deg2-33-pre1}
The branching vectors of Rule~(\ref{rule:deg2-33-pre1}) are dominated by
$(1+\Psi(1,1),1+\Psi(1,1),3)$.
\end{lemma}
\begin{proof}
Let $G_1 = \instance{G}{+x,-y}$, $G_2 = \instance{G}{-x,+y}$, and $G_3 = \instance{G}{-x,-y}$.
Let $z_i$ be the unique vertex in $e_i \setminus \{x,y\}$.
Since $y$ is not dominated by $x$ there is a hyperedge $e$ that contains $y$ and do not contain $x$.
The hyperedge $e$ becomes a 2-hyperedge in the hypergraph $G_1$.
Therefore, $m_2(G_1) \geq 1$.
Similarly, $m_2(G_2) \geq 1$.
The hyperedges $e_1,e_2,e_3$ become 1-hyperedges in $G_3$.
Therefore, the algorithm applies Rule~(\ref{rule:loop}) three times starting from
the instance $(G_3,k)$.
We assume that these application occur inside Rule~(\ref{rule:deg2-33-pre1}).
Namely, we assume that the rule generates the instances
$(G_1,k-1)$, $(G_2,k-1)$, and $(G'_3,k-3)$, where
$G'_3 = \instance{G}{+z_1,+z_2,+z_3,-x,-y}$.
Therefore, the branching vectors of Rule~(\ref{rule:deg2-33-pre1}) are dominated by
$(1+\Psi(1,1),1+\Psi(1,1),3)$.
\end{proof}

\begin{lemma}\label{lem:rule-deg2-33-pre2}
The branching vectors of Rule~(\ref{rule:deg2-33-pre2}) are dominated by
$(1+\Psi(4,*),2,2)$.
\end{lemma}
\begin{proof}
By Claim~\ref{clm:deg-2-intersection}, $e_1 \cap e_2 = \{x\}$.
Let $G_1 = \instance{G}{+x,-y,-z,-v,-w}$.
Since $I(x) = 0$, then $|e \cap N(x)| = 1$ for every hyperedge $e \in E(G) \setminus \{e_1,e_2\}$
such that $e \cap N(x) \neq \emptyset$.
Therefore, $\add{G}{G_1} = B$.
By the condition of the rule we have $m_2(G_1) = |B| = 4$.
and the lemma follows.
\end{proof}

\begin{lemma}\label{lem:rule-deg2-33}
The branching vectors of Rule~(\ref{rule:deg2-33}) are dominated by
\begin{align*}
&\{           (1+\Psi(5,*),\Psi(2,2)), (1+\Psi(4,*),\min(2,\Psi(4,2))),\\
&\phantom{\{} (1+\Psi(4,*),1+\Psi(3,*),1+\Psi(2,2)), (1+\Psi(4,*),2,1+\Psi(2,*)),\\
&\phantom{\{} (\min(3,2+\Psi(1,1)),\Psi(2,2))\}.
\end{align*}
\end{lemma}
\begin{proof}
By Claim~\ref{clm:deg-2-intersection} we have $e_1 \cap e_2 = \{x\}$.
Rule~(\ref{rule:deg2-33}) generates the instances
$(G_1 = \instance{G}{+x,-y,-z,-v,-w}, k-1)$ and $(G_2 = \instance{G}{-x},k)$.
Since $m_2(G) = 0$ (Claim~\ref{clm:m2-0}) and $\add{G}{G_2} = \{e_1,e_2\}$,
then $m_2(G_2) = 2$.
We also have $d_2(G_2) = 1$ and $c_2(G_2) = 2$ since $e_1 \cap e_2 = \{x\}$.

Assume that $I(x) = 0$.
If $|\add{G}{G_1}| \geq 5$, then the branching vector for the instance $(G,k)$ is dominated by
$(1+\Psi(5,*),\Psi(2,2))$.
Now suppose that $|\add{G}{G_1}| \leq 4$.
Let $B$ the set and $H$ be the bipartite graph that are defined in Rule~(\ref{rule:deg2-33-pre2}).
Since $I(x) = 0$, then $|e \cap N(x)| = 1$ for every hyperedge $e \in E(G) \setminus \{e_1,e_2\}$
such that $e \cap N(x) \neq \emptyset$.
Therefore, $\add{G}{G_1} = B$.
By Claim~\ref{clm:min-deg-2} we have $d(p) \geq 2$ for every $p \in N(x)$.
Let $N_2 = \{p : p \in N(x), d(p) = 2\}$.

For $p \in N_2$, let $e_p$ be the unique hyperedge in $E(G) \setminus \{e_1,e_2\}$ that contains $p$.
For $p \in N_2$ we have that $e_p \setminus N(x) \neq e \setminus N(x)$ for every $e \neq e_p$,
otherwise we have $I(p) = 1$, a contradiction to the choice of $x$ in Rule~(\ref{rule:deg2-33}).
Therefore, $d^H(e_p \setminus N(x)) = 1$ for every $p \in N_2$.
By Claim~\ref{clm:xy}, $d^H(e) \leq 2$ for every $e \in B$. Therefore,
$|E_H| = \sum_{e \in B} d^H(e) \leq |N_2| + 2|B \setminus N_2| = 2|B|-|N_2|$.
We also have
$|E_H| = \sum_{p \in N(x)} d^H(p) = |N_2| + \sum_{p \in N(x) \setminus N_2} d^H(p)
\geq |N_2|+2|N(x) \setminus N_2| = 8-|N_2|$.
Therefore, $|B| \geq 4$.
Since we assumed that $|B| \leq 4$, we obtain that
$|B| = |\add{G}{G_1}| = 4$.
Additionally, $d^H(p) = 2$ for every $p \in N(x) \setminus N_2$,
and therefore $d(p) = 3$ for every $p \in N(x) \setminus N_2$.
If $N_2 = \emptyset$, then $x$ satisfies the conditions of Rule~(\ref{rule:deg2-33-pre2}),
a contradiction.
Therefore, $N_2 \neq \emptyset$.

Let $p \in N_2$.
We have that $d^{G_2}(p) = 2$ and $d_2^{G_2}(p) = 1$
(the hyperedges of $G_2$ containing $p$ are $e_1 \setminus \{x\}$ or $e_2 \setminus \{x\}$ and $e_p$).
Therefore, $p$ satisfies the condition of Rule~(\ref{rule:deg2-2x}) in the hypergraph $G_2$,
so the algorithm applies Rule~(\ref{rule:deg2-2x}) on $(G_2,k)$ or an earlier rule.
Since every hyperedge in $G$ has size~3 (by Claim~\ref{clm:m2-0}),
then every hyperedge in $G_2$ has size at least 2
(recall that $|e \cap N(x)| = 1$ for every hyperedge $e \in E(G) \setminus \{e_1,e_2\}$
such that $e \cap N(x) \neq \emptyset$).
Therefore,  Rule~(\ref{rule:loop}) cannot be applied on $(G_2,k)$.
The only 2-hyperedges in $G_2$ are $\{y,z\}$ and $\{v,w\}$.
Therefore, Rule~(\ref{rule:cycle}) cannot be applied on $(G_2,k)$.
No 3-hyperedge of $G_2$ contains $\{y,z\}$ or $\{v,w\}$.
Therefore, Rule~(\ref{rule:non-simple}) cannot be applied on $(G_2,k)$.
If $x'$ is a vertex that is dominated by a vertex $y$ in the hypergraph $G_2$, then
$x'$ is also dominated by $y$ in $G$, a contradiction to the fact that Rule~(\ref{rule:dominated})
cannot be applied on $(G,k)$.
Thus, Rule~(\ref{rule:dominated}) cannot be applied on $(G_2,k)$.
Therefore, the algorithm applies a rule from Rules~(\ref{rule:cc}),
(\ref{rule:deg2-c}), (\ref{rule:deg2-u}), and (\ref{rule:deg2-2x}) on $(G_2,k)$.
We assume that this application occurs inside Rule~(\ref{rule:deg2-33}).

Suppose that Rule~(\ref{rule:cc}) or Rule~(\ref{rule:deg2-u}) is applied on $(G_2,k)$.
These rules decrease the parameter $k$ by at least one.
Therefore, the branching vector of Rule~(\ref{rule:deg2-33}) for the instance $(G,k)$ is dominated by
$(1+\Psi(4,*),1)$.

Suppose that Rule~(\ref{rule:deg2-c}) is applied on $(G_2,k)$.
This rule returns the instance $(G'_2,k)$, where $G'_2 = \instance{G_2}{-x'} = \instance{G}{-x,-x'}$
for some vertex $x'$.
We assume that Rule~(\ref{rule:deg2-33}) returns the instances $(G_1,k-1)$ and $(G'_2,k)$.
By the condition of Rule~(\ref{rule:deg2-c}) we have that there is a hyperedge $e \in E(G_2)$
such that $e \subseteq N^{G_2}(x')$.
We have that $d(x') = d^{G_2}(x') = 2$ and $e \subseteq N^{G_2}(x') \subseteq N(x')$.
Since Rule~(\ref{rule:deg2-c}) cannot be applied on $(G,k)$, it follows that
$e \notin E(G)$.
By the definition of $G_2$, we have that either $e = \{y,z\}$ or $e = \{v,w\}$.
Without loss of generality, $e = \{y,z\}$.
Let $e_3,e_4$ be the hyperedges of $G_2$ containing $x'$.
Since $I(x) = 0$, then $|e_3 \cap N(x)| \leq 1$ and $|e_4 \cap N(x)| \leq 1$
($e_3$ and $e_4$ are either hyperedges in $G$ or subsets of hyperedges in $G$).
Therefore, $e \not\subseteq e_3$ and $e \not\subseteq e_4$.
Recall that $e \subseteq N^{G_2}(x') = (e_3 \setminus \{x'\}) \cup (e_4 \setminus \{x'\})$.
It follows that $e_3 \setminus \{x'\}$ contains one vertex of $e$ and
$e_4 \setminus \{x'\}$ contains the other vertex of $e$.
Without loss of generality, $y \in e_3$ and $z \in e_4$.
Since $|e_3 \cap N(x)| \leq 1$ we obtain that $e_3 \neq \{y,z\}$ and $e_3 \neq \{v,w\}$,
and thus $|e_3| = 3$.
Since $|e_3 \cap N(x)| \leq 1$ and $y,x' \in e_3$, then $x' \notin N(x)$.
Since $|e_4 \cap N(x)| \leq 1$, then $|e_4| = 3$.
The graph $G'_2$ has exactly four 2-hyperedges: $\{v,w\}$, $\{y,z\}$,
$e_3 \setminus \{y\}$, and $e_4 \setminus \{z\}$.
The last three hyperedges are in the same connected component of $(G'_2)^2$, and
the hyperedge $\{v,w\}$ is in a different connected component.
Namely, we have $c_2(G'_2) = 2$.
Therefore, the branching vector of Rule~(\ref{rule:deg2-33}) for the instance $(G,k)$ is dominated by
$(1+\Psi(4,*),\Psi(4,2))$.

Suppose that Rule~(\ref{rule:deg2-2x}) is applied on $(G_2,k)$ and generates instances
$(H_1,k_1)$ and $(H_2,k_2)$.
We assume that Rule~(\ref{rule:deg2-33}) returns the instances $(G_1,k-1)$, $(H_1,k_1)$, and $(H_2,k_2)$.
By Lemma~\ref{lem:rule-deg2-2x-1}, either
\begin{align*}
\mu(G_2,k)-\mu(H_1,k_1) & \geq 1-\Delta(2,-1,2,c') = 1-\Psi(2,2)+\Psi(3,c') \geq 1-\Psi(2,2)+\Psi(3,*) \\
\mu(G_2,k)-\mu(H_2,k_2) & \geq 1
\end{align*}
or
\begin{align*}
\mu(G_2,k) - \mu(H_1,k_1) & \geq 2-\Delta(2,3,2,0) = 2-\Psi(2,2)+\Psi(0,0) = 2-\Psi(2,2) \\
\mu(G_2,k) - \mu(H_2,k_2) & \geq 1-\Delta(2,0,2,c') = 1-\Psi(2,2)+\Psi(2,c') \geq 1-\Psi(2,2)+\Psi(2,*).
\end{align*}
In the first case we have
\begin{align*}
\mu(G,k)-\mu(H_1,k_1)
 & = (\mu(G,k)-\mu(G_2,k)) + (\mu(G_2,k)-\mu(H_1,k_1))\\
 & \geq \Psi(2,2) + (1-\Psi(2,2)+\Psi(3,*))\\
 & = 1+\Psi(3,*)
\end{align*}
and
\[ \mu(G,k)-\mu(H_2,k_2) = (\mu(G,k)-\mu(G_2,k)) + (\mu(G_2,k)-\mu(H_2,k_2)) \geq \Psi(2,2) + 1. \]
Therefore, the branching vector of Rule~(\ref{rule:deg2-33}) for the instance $(G,k)$
is dominated by $(1+\Psi(4,*),1+\Psi(3,*),1+\Psi(2,2))$.
Similarly, in the second case, the branching vector of Rule~(\ref{rule:deg2-33}) for the instance $(G,k)$
is dominated by $(1+\Psi(4,*),2,1+\Psi(2,*))\}$.

We now consider the case when $I(x) = 1$.
By definition, there is a hyperedge $e$ such that $x \notin e$ and $|e \cap N(x)| \geq 2$.
Since Rule~(\ref{rule:deg2-c}) cannot be applied on $(G,k)$,
$e$ contains a vertex $u \notin N(x)$.
Since $\{u\}$ is a 1-hyperedge in $G_1$, the algorithm applies Rule~(\ref{rule:loop})
on $(G_1,k-1)$ which generates the instance $(G'_1 = \instance{G}{+x,+u,-y,-z,-v,-w},k-2)$.
We assume that Rule~(\ref{rule:deg2-33}) generates the instances $(G'_1,k-2)$ and $(G_2,k)$.
Since Rule~(\ref{rule:deg2-u}) cannot be applied on $(G,k)$,
then there is a hyperedge $e'$ such that $e' \cap N(x) \neq \emptyset$ and $x,u \notin e'$.
If $|e' \cap N(x)| = 1$, then $e'$ becomes a 2-hyperedge in $G'_1$.
Therefore, the branching vector for the instance $(G,k)$ is dominated by
$(2+\Psi(1,1),\Psi(2,2))$.
Otherwise, we have $|e' \cap N(x)| = 2$.
Let $u'$ be the unique vertex in $e' \setminus N(x)$.
The hyperedge $e'$ becomes a 1-hyperedge in $G'_1$, and therefore
the algorithm applies Rule~(\ref{rule:loop})
on $(G'_1,k-1)$ which generates the instance $(G''_1 = \instance{G}{+x,+u,+u',-y,-z,-v,-w},k-3)$.
Therefore, the branching vector for the instance $(G,k)$ is dominated by
$(3,\Psi(2,2))$.
\end{proof}

\subsection{Analysis of Rule~(\ref{rule:final})}
\label{sec:analysis-rule-final}

In this section we analyzes Rule~(\ref{rule:final}).
Let $d = d(G)$.
Due to Lemma~\ref{lem:path}, we only need to consider the case $d \geq 4$.

We first consider the case $d \geq 6$.
We have that $m_2(G) = 0$ (by Claim~\ref{clm:m2-0}) and therefore $m_2(\instance{G}{+u}) = 0$.
Additionally, $\add{G}{\instance{G}{-u}}$ consists of the $d$ hyperedges that contain $u$,
and therefore $m_2(\instance{G}{-u}) = d \geq 6$.
It follows that the branching vectors of Rule~(\ref{rule:final})
in this case are dominated by $(1,\Psi(6,*))$.

We now assume that $d \in \{4,5\}$.
Rule~(\ref{rule:final}) generates the instances $(G_1 = \instance{G}{+u},k-1)$
and $(\instance{G}{-u},k)$.
If $\instance{G}{-u}$ if simple, then let $G_2 = \instance{G}{-u}$.
Otherwise, let $G_2$ be the hypergraph obtained by applying Rule~(\ref{rule:non-simple})
on $(\instance{G}{-u},k)$.
Note that in the latter case, Rule~(\ref{rule:non-simple})
deletes only 3-hyperedges from $\instance{G}{-u}$ (since all the hyperedges in $G$ have size~3
by Claim~\ref{clm:m2-0} and therefore every hyperedge in $\instance{G}{-u}$ has size at least~2)
and therefore $m_2(G_2) = m_2(\instance{G}{-u})$.
In both cases, $G_2$ does not contain 1-hyperedges and $G_2$ is a simple hypergraph.

Since $m_2(G_2) = d > 0$, the algorithm applies Rule~(\ref{rule:d2}) on $(G_2,k)$
or an earlier rule.
Since Rules~(\ref{rule:loop}) and~(\ref{rule:non-simple}) cannot be applied on $(G_2,k)$,
the algorithm applies on $(G_2,k)$ a rule from Rules~(\ref{rule:cc})--(\ref{rule:d2}).
In the analysis below, we assume that this application occurs inside the application
of Rule~(\ref{rule:final}) on $(G,k)$.

We first assume that Rule~(\ref{rule:cc}) or Rule~(\ref{rule:deg2-u}) is applied on $(G_2,k)$.
These rules decrease the parameter $k$ by at least 1.
Therefore, the branching vectors of Rule~(\ref{rule:final}) in this case are dominated by $(1,1)$.

Now suppose that Rule~(\ref{rule:dominated}) or Rule~(\ref{rule:deg2-c}) is applied on $(G_2,k)$,
and let $(G'_2,k)$ be the resulting instance.
By the proof of Lemma~\ref{lem:reduction},
the application of Rule~(\ref{rule:dominated}) or Rule~(\ref{rule:deg2-c})
either increases the number of 2-hyperedges or creates at least one 1-hyperedge.
In the former case we have that $m_2(G'_2) \geq d+1$,
so the branching vectors in this case are dominated by $(1,\Psi(d+1,*))$.
In the latter case we have that Rule~(\ref{rule:loop}) is applied on $(G'_2,k)$,
and we assume that Rule~(\ref{rule:final}) generates the instances $(G_1,k-1)$ and $(G''_2,k-1)$,
where $(G''_2,k-1)$ is the instance obtained from the application of Rule~(\ref{rule:loop}) on $(G'_2,k)$.
Thus, the branching vectors of Rule~(\ref{rule:final}) in this case are dominated by $(1,1)$.

Suppose that Rule~(\ref{rule:cycle}) is applied on $(G_2,k)$,
and let $(H_1,k_1)$ and $(H_2,k_2)$ be the instances generated by this rule.
We assume that the instances generated by Rule~(\ref{rule:final}) are
$(G_1,k-1)$, $(H_1,k_1)$, and $(H_2,k_2)$.
We have $k_i = k-2$.
Therefore, the branching vectors of Rule~(\ref{rule:final}) in this case are dominated by $(1,2,2)$.

Suppose that Rule~(\ref{rule:deg2-2x}) is applied on $(G_2,k)$,
and let $(H_1,k_1)$ and $(H_2,k_2)$ be the instances generated by the rule.
If the first case of Rule~(\ref{rule:deg2-2x}) is applied, then by By Lemma~\ref{lem:rule-deg2-2x-2} either
\begin{align*}
\mu(G_2,k)-\mu(H_1,k_1) & \geq 1-\Delta(d,1,c,c') = 1-\Psi(d,c)+\Psi(d-1,c')\\
                        & \geq 1-\Psi(d,c)+\Psi(d-1,*) \\
\mu(G_2,k)-\mu(H_2,k_2) & \geq 2-\Delta(d,2,c,c-1) = 2-\Psi(d,c)+\Psi(d-2,c-1)\\
                        & \geq 2-\Psi(d,c)+\Psi(d-2,*)
\end{align*}
or
\begin{align*}
\mu(G_2,k)-\mu(H_1,k_1) & \geq 2-\Delta(d,5,c,c') = 2-\Psi(d,c) \\
\mu(G_2,k)-\mu(H_2,k_2) & \geq 2-\Delta(d,4,c,c'') = 2-\Psi(d,c)+\Psi(d-4,c'')\\
                        & \geq 2-\Psi(d,c)+\Psi(d-4,*).
\end{align*}
where $c = c_2(G_2)$.
In the first case, we have
\begin{align*}
\mu(G,k)-\mu(H_1,k_1)
 & = (\mu(G,k)-\mu(G_2,k)) + (\mu(G_2,k)-\mu(H_1,k_1))\\
 & \geq \Psi(d,c) + (1-\Psi(d,c)+\Psi(d-1,*))\\
 & = 1+\Psi(d-1,*)
\end{align*}
and
\begin{align*}
\mu(G,k)-\mu(H_2,k_2)
 & = (\mu(G,k)-\mu(G_2,k)) + (\mu(G_2,k)-\mu(H_2,k_2))\\
 & \geq \Psi(d,c) + (2-\Psi(d,c)+\Psi(d-2,*))\\
 & = 2+\Psi(d-2,*)
\end{align*}
Therefore, the branching vectors of Rule~(\ref{rule:final}) in this case are dominated by
$(1,1+\Psi(d-1,*),2+\Psi(d-2,*))$.
Similarly, in the second case, the branching vectors of Rule~(\ref{rule:final}) are dominated by
$(1,2,2+\Psi(d-4,*))$.
If the second case of Rule~(\ref{rule:deg2-2x}) is applied, then by Lemma~\ref{lem:rule-deg2-2x-1},
the branching vectors of Rule~(\ref{rule:final}) are dominated by
\[ \{(1,1+\Psi(d+1,*),1+\Psi(d,d)), (1,2+\Psi(d-3,*),1+\Psi(d,*))\}. \]

Suppose that Rule~(\ref{rule:d2}) is applied on $(G_2,k)$.
Recall that the rule is applied on a vertex $x$.
Let $d' = d_2(G_2) = d_2^{G_2}(x)$.
Let $\{x,y_1\},\ldots,\{x,y_{d'}\}$ be the 2-hyperedges of $G_2$ that contain $x$.
We assume that the instances that are generated by Rule~(\ref{rule:d2}) are
$(H_1,k-1)$, and $(H_2,k-d')$,
where $H_1 = \instance{G_2}{+x}$ and $H_2 = \instance{G_2}{-x,+y_1,\ldots,+y_{d'}}$
(see the proof of Lemma~\ref{lem:rule-d2}).
We assume that Rule~(\ref{rule:final}) generates the instances
$(G_1,k-1)$, $(H_1,k-1)$, and $(H_2,k-d')$.
Let $e^u_1,\ldots,e^u_d$ be the hyperedges of $G$ that contain $u$.
The 2-hyperedges of $G_2$ are $e^u_1 \setminus \{u\}, \ldots, e^u_d \setminus \{u\}$.
Since $u$ is not dominated by $x$ in $G$ (as Rule~(\ref{rule:dominated}) cannot be applied on $(G,k)$),
then at least one of the hyperedges $e_1 \setminus \{u\}, \ldots, e_d \setminus \{u\}$ does not contain $x$.
Therefore, $d' < d $.

Suppose that $d = 5$.
By Lemma~\ref{lem:rule-d2}, if $d' = 1$ then
the branching vectors of Rule~(\ref{rule:final}) in this case are dominated by
$(1,1+\Psi(4,4),1+\Psi(6,*))$.
Additionally, if $d' = 2$ then
the branching vectors of Rule~(\ref{rule:final}) in this case are dominated by
$(1,1+\Psi(3,*),2+\Psi(2,*))$,
and if $d' \in \{3,4\}$ then
the branching vectors are dominated by
$(1,1+\Psi(5-d',*),d')$.

We now consider the case $d = 4$. For this case we need stronger bounds on the branching vectors
of the rule.

\begin{observation}\label{obs:d-G1}
For every $v \in N(u)$, $d^{G_1}(v) = d(v)-d_2^{G_2}(v)$.
For every $v \in V(G_1) \setminus N(u)$, $d^{G_1}(v) = d(v)$.
\end{observation}

Since Rule~(\ref{rule:deg2-33-pre1}) cannot be applied on $(G,k)$,
we have that $d' \neq 3$.
Therefore, either $d' = 1$ or $d' = 2$.

\subsubsection{$d' = 1$}
Suppose that $d' = 1$.
We have that $d(G) = 4$ and $d_2^{G_2}(x) = 1$, and therefore
$d_3^{G_2}(x) = d^{G_2}(x)-d_2^{G_2(x)} \leq d(x)-1 \leq 3$.
Since $d_2^{G_2}(x) = 1 = d_2(G_2)$,
by Claim~\ref{clm:dv} (on the instance $(G_2,k)$) we have that
$d_3^{G_2}(x) = d^{G_2}(x)-d_2^{G_2(x)} \geq 3-1$.
Therefore, $d_3^{G_2}(x) \in \{2,3\}$.
If $d_3^{G_2}(x) = 3$, then by Lemma~\ref{lem:rule-d2} we have that
the branching vectors of Rule~(\ref{rule:final}) in this case are dominated by
$(1,1+\Psi(3,3),1+\Psi(6,*))$.
Assume for the rest of the section that $d_3^{G_2}(x) = 2$.

\begin{lemma}\label{lem:G2-1}
$G_2 = \instance{G}{-u}$.
\end{lemma}
\begin{proof}
Suppose for contradiction that $G_2 \neq \instance{G}{-u}$.
By the definition of $G_2$ we have that there are hyperedges $e,e' \in E(\instance{G}{-u})$
such that $e' \subset e$.
By the definition of $\instance{G}{-u}$ we have that $e' = \{v,w\}$ for some vertices
$v,w$ such that $\{u,v,w\}$ is a hyperedge of $G$.
Since $e \in E(\instance{G}{-u}) \setminus E(G_2)$, then
$d(v) = d^{\instance{G}{-u}}(v) \geq d^{G_2}(v) + 1$.
Since $d_2^{G_2}(v) = 1 = d_2(G_2)$,
then by Claim~\ref{clm:dv} we have $d^{G_2}(v) \geq 3$ and therefore $d(v) \geq 4$.
Since $d(G) = 4$, then $d(v) = 4 = d(u)$.
We have $d_2(\instance{G}{-u}) = d' = 1$ and
$d_2(\instance{G}{-v}) \geq 2$
(due to the 3-hyperedges $\{u,v,w\}$ and $e$ that become intersecting 2-hyperedges in $\instance{G}{-v}$),
a contradiction to the choice of $u$ in Rule~(\ref{rule:final}).
\end{proof}

\begin{lemma}\label{lem:Nx}
$d^{G_1}(v) = 2$ for every $v \in N(u)$.
\end{lemma}
\begin{proof}
Fix $v \in N(u)$.
Since $d_2^{G_2}(v) = 1 = d_2(G_2)$,
then by Claim~\ref{clm:dv}, $d^{G_2}(v) \geq 3$.
By the choice of $x$ in Rule~(\ref{rule:d2}) and the fact that
$d_2^{G_2}(v) = d_2(G_2)$
we have that
$d_3^{G_2}(x)-D_2^{G_2}(x) \geq d_3^{G_2}(v)-D_2^{G_2}(v)$.
Since $D_2^{G_2}(v) = D_2^{G_2}(x) = 1$,
we obtain that $d_3^{G_2}(v) \leq d_3^{G_2}(x) = 2$.
Therefore, $d^{G_2}(v) = d_3^{G_2}(v)+1 \leq 3$.
Thus, $d^{G_2}(v) = 3$.
By Lemma~\ref{lem:G2-1}, $d(v) = d^{\instance{G}{-u}}(v) = d^{G_2}(v) = 3$.
By Observation~\ref{obs:d-G1}, $d^{G_1}(v) = 2$.
\end{proof}

By Lemma~\ref{lem:Nx}, the algorithm applies Rule~(\ref{rule:deg2-33}) on $(G_1,k-1)$ 
or an earlier rule.
Since all the hyperedges of $G_1$ have size~3 (by Claim~\ref{clm:m2-0}) then the rules that the algorithm
can apply on $(G_1,k-1)$ are (\ref{rule:cc}), (\ref{rule:dominated}), (\ref{rule:deg2-c}),
(\ref{rule:deg2-u}), (\ref{rule:deg2-33-pre1}), (\ref{rule:deg2-33-pre2}), and~(\ref{rule:deg2-33}).

If the algorithm applies Rule~(\ref{rule:cc}) or Rule~(\ref{rule:deg2-u}) on $(G_1,k-1)$,
then the value of the parameter $k$ is decreased by at least 1.
Therefore, the branching vectors of Rule~(\ref{rule:final}) in this case are dominated by
$(2,1+\Psi(3,3),1+\Psi(5,*))$.

If the algorithm applies Rule~(\ref{rule:dominated}) or Rule~(\ref{rule:deg2-c}) on $(G_1,k-1)$,
then the resulting instance is $(G'_1 = G_1[-x'], k-1)$ for some vertex $x'$.
Since $m_2(G_1) = 0$, then the 2-hyperedges of $G'_1$ are the hyperedges of $G_1$ that contain $x'$,
after removing $x'$ from these hyperedges.
Thus, $m_2(G'_1) = d^{G_1}(x')$.
We have $d^{G_1}(x') = 2$:
If the algorithm applies Rule~(\ref{rule:deg2-c}) on $(G_1,k-1)$,
then $d^{G_1}(x') = 2$ by the definition of Rule~(\ref{rule:deg2-c}).
If the algorithm applies Rule~(\ref{rule:dominated}) on $(G_1,k-1)$,
then $x'$ is dominated by another vertex $y$.
Since $x'$ is not dominated in $G$, there is a hyperedge in $E(G) \setminus E(G_1)$
that contains $x'$ and does not contain $y$.
By the definition of $G_1$ we have $x' \in N(u)$.
Therefore, by Lemma~\ref{lem:Nx}, $d^{G_1}(x') = 2$.
We showed above that $m_2(G'_1) = 2$.
We assume that Rule~(\ref{rule:final}) generates the instances
$(G'_1,k-1)$, $(H_1,k-1)$, and $(H_2,k-2)$.
Therefore, the branching vectors of Rule~(\ref{rule:final}) in this case are dominated by
$(1+\Psi(2,*),1+\Psi(3,3),1+\Psi(5,*))$.

If the algorithm applies Rule~(\ref{rule:deg2-33-pre1}) on $(G_1,k-1)$,
then by Lemma~\ref{lem:rule-deg2-33-pre1},
the branching vectors of Rule~(\ref{rule:final}) in this case are dominated by
$(2+\Psi(1,1),2+\Psi(1,1),4,1+\Psi(3,3),1+\Psi(5,*))$.
If the algorithm applies Rule~(\ref{rule:deg2-33-pre2}) on $(G_1,k-1)$,
then by Lemma~\ref{lem:rule-deg2-33-pre2},
the branching vectors of Rule~(\ref{rule:final}) in this case are dominated by
$(2+\Psi(4,*),3,3,1+\Psi(3,3),1+\Psi(5,*))$.
If the algorithm applies Rule~(\ref{rule:deg2-33}) on $(G_1,k-1)$,
then by Lemma~\ref{lem:rule-deg2-33},
the branching vectors of Rule~(\ref{rule:final}) in this case are dominated by
\begin{align*}
& \{(2+\Psi(5,*),1+\Psi(2,2),1+\Psi(3,3),1+\Psi(5,*)),\\
&\phantom{\{} (2+\Psi(4,*),\min(2,1+\Psi(4,2)),1+\Psi(3,3),1+\Psi(5,*)),\\
&\phantom{\{} (2+\Psi(4,*),2+\Psi(3,*),2+\Psi(2,2),1+\Psi(3,3),1+\Psi(5,*)),\\
&\phantom{\{} (2+\Psi(4,*),3,2+\Psi(2,*),1+\Psi(3,3),1+\Psi(5,*)),\\
&\phantom{\{} (\min(4,3+\Psi(1,1)),1+\Psi(2,2),1+\Psi(3,3),1+\Psi(5,*))\}.
\end{align*}

\subsubsection{$d' = 2$}
We now consider the case $d' = 2$.
Since $d(x) \geq d^{G_2}(x) \geq 3$ (by Claim~\ref{clm:dv})
and $d(G) = 4$, then $d(x) \in \{3,4\}$.
By Observation~\ref{obs:d-G1}, $d^{G_1}(x) = d(x)-2 \in \{1,2\}$.
Therefore, the algorithm applies Rule~(\ref{rule:deg2-33}) on $(G_1,k-1)$ or an earlier rule.
Since all the hyperedges in $G_1$ have size~3, then the algorithm applies a rule from
(\ref{rule:cc}), (\ref{rule:dominated}), (\ref{rule:deg2-c}), (\ref{rule:deg2-u})
(\ref{rule:deg2-33-pre1}), (\ref{rule:deg2-33-pre2}), and (\ref{rule:deg2-33}).

If the algorithm applies Rule~(\ref{rule:cc}) or Rule~(\ref{rule:deg2-u}) on $(G_1,k-1)$, then the parameter $k$
decreases by at least 1, and by Lemma~\ref{lem:rule-d2},
the branching vectors of Rule~(\ref{rule:final}) in this case are dominated by
$(2,1+\Psi(2,*),1+\Psi(2,*))$.
If the algorithm applies Rule~(\ref{rule:dominated}) or Rule~(\ref{rule:deg2-c}) on $(G_1,k-1)$, then
the resulting instance is $(G'_1 = G_1[-x'], k-1)$ for some vertex $x'$.
Since $m_2(G_1) = 0$, then $m_2(G'_1) = d^{G_1}(x') \geq 1$.
Therefore, the branching vectors of Rule~(\ref{rule:final}) in this case are dominated by
$(1+\Psi(1,1),1+\Psi(2,*),2+\Psi(2,*))$.
If the algorithm applies Rule~(\ref{rule:deg2-33-pre1}) on $(G_1,k-1)$, then
by Lemma~\ref{lem:rule-deg2-33-pre1},
the branching vectors of Rule~(\ref{rule:final}) in this case are dominated by
$(2+\Psi(1,1),2+\Psi(1,1),4,1+\Psi(2,*),2+\Psi(2,*))$.
If the algorithm applies Rule~(\ref{rule:deg2-33-pre2}) on $(G_1,k-1)$, then
by Lemma~\ref{lem:rule-deg2-33-pre2},
the branching vectors of Rule~(\ref{rule:final}) in this case are dominated by
$(2+\Psi(4,*),3,3,1+\Psi(2,*),2+\Psi(2,*))$.
If the algorithm applies Rule~(\ref{rule:deg2-33}) on $(G_1,k-1)$, then
by Lemma~\ref{lem:rule-deg2-33},
the branching vectors of Rule~(\ref{rule:final}) in this case are dominated by
\begin{align*}
& \{(2+\Psi(5,*),1+\Psi(2,2),1+\Psi(2,*),2+\Psi(2,*)),\\
&\phantom{\{} (2+\Psi(4,*),\min(2,1+\Psi(4,2)),1+\Psi(2,*),2+\Psi(2,*)),\\
&\phantom{\{} (2+\Psi(4,*),2+\Psi(3,*),2+\Psi(2,2),1+\Psi(2,*),2+\Psi(2,*)),\\
&\phantom{\{} (2+\Psi(4,*),3,2+\Psi(2,*),1+\Psi(2,*),2+\Psi(2,*)),\\
&\phantom{\{} (\min(4,3+\Psi(1,1)),1+\Psi(2,2),1+\Psi(2,*),2+\Psi(2,*))\}.
\end{align*}

\subsection{Analysis of the algorithm}

To obtain the functions $\Psi_3,\Psi_4,\Psi_5,\Psi_6$
we used convex programming (see~\cite{gaspers2008exponential}).
The function $\Psi_i$ is obtained by creating a convex program
whose variables are the values of $\Psi_i(m,c)$ for every $m \in [1,8]$ and $c \in [1,m]$
(for every $m' > 8$ and $c \in [1,m']$ we define $\Psi_i(m',c) = \Psi_i(8,1)$).
The goal of the program is to minimize the maximum branching number of the rules
in the case when $\hat{d}(G) = i$.
The relevant branching vectors in Sections~\ref{sec:analysis-branching-rules}
and~\ref{sec:analysis-rule-final} define the constraints of the program.
An AMPL code that defines and solves this convex program is available at
\url{https://github.com/dekelts/hs}.
The function $\Psi_4$ is shown in Table~\ref{tab:psi4}.
For the functions $\Psi_3,\Psi_4,\Psi_5,\Psi_6$ constructed by this process,
the branching numbers of the branching rules of the algorithm are given in Table~\ref{tab:bn}.

\begin{table}
\centering
\begin{tabular}{clllllllll}
\diagbox[width=30pt, height=20pt]{$m$}{$c$}
 & 0 & 1 & 2 & 3 & 4 & 5 & 6 & 7 & $\geq 8$ \\
\midrule
0 & 0 & 0 & 0 & 0 & 0 & 0 & 0 & 0 & 0 \\
1 & 0 & 0.244 & 0 & 0 & 0 & 0 & 0 & 0 & 0 \\
2 & 0 & 0.5154 & 0.4706 & 0 & 0 & 0 & 0 & 0 & 0 \\
3 & 0 & 0.6842 & 0.6842 & 0.6733 & 0 & 0 & 0 & 0 & 0 \\
4 & 0 & 0.8441 & 0.8898 & 0.9087 & 0.8742 & 0 & 0 & 0 & 0 \\
5 & 0 & 1.0444 & 1.0444 & 1.0444 & 1.0444 & 1.0597 & 0 & 0 & 0 \\
6 & 0 & 1.2109 & 1.2108 & 1.2108 & 1.2109 & 1.214 & 1.2088 & 0 & 0 \\
7 & 0 & 1.3151 & 1.315 & 1.315 & 1.315 & 1.3152 & 1.316 & 1.3129 & 0 \\
$\geq 8$ & 0 & 1.3666 & 1.3666 & 1.3666 & 1.3666 & 1.3666 & 1.3666 & 1.3666 & 1.3666 \\
\end{tabular}
\caption{The function $\Psi_4$.\label{tab:psi4}}
\end{table}

\begin{table}
\centering
\begin{tabular}{clllllll}
$\hat{d}(G)$ & B1 & B2 & B3 & B4 & B5 & B6 & B8\\
\midrule
3 & 1.816 & 2.0 & 2.0299 & NA & 1.7709 & 2.0299 & - \\
4 & 1.8215 & 2.0 & 2.0409 & 1.9584 & 1.7585 & 1.9423 & 2.0409 \\
5 & 1.812 & 2.0 & 2.0345 & NA & NA & NA & 2.0345 \\
6 & 1.7839 & 2.0 & 2.032 & NA & NA & NA & 2.032 \\
\end{tabular}
\caption{The branching numbers of the branching rules
for different values of $\hat{d}(G)$.
NA indicates that the rule cannot be applied due to a condition $d(G) = 4$ or $d(G) \leq 4$
in the rule.
We do not bound the branching number of Rule~(\ref{rule:final})
in the case $\hat{d}(G) \leq 3$ due to Lemma~\ref{lem:path}.\label{tab:bn}}
\end{table}

Let $L(G,k)$ be the number of leaves in the recursion tree for $(G,k)$.
Let $\alpha(G,k) = 3$ if the recursion tree for $(G,k)$ contains a node that corresponds
to an application of rule~(\ref{rule:final}) on an instance $(G',k')$ with $d(G') \leq 3$.
Otherwise, $\alpha(G,k) = 0$.

\begin{lemma}\label{lem:mu}
Let $(G,k)$ be an instance and let $(G',k')$ be an instance that is obtained by
applying a reduction rule or a branching rule on $(G,k)$.
Then,
$\mu(G',k')+2\hat{d}(G') \leq \mu_{\hat{d}(G)}(G',k')+2\hat{d}(G)$.
\end{lemma}
\begin{proof}
$G'$ is obtained from $G$ by deleting vertices and hyperedges.
Therefore, $\hat{d}(G') \leq \hat{d}(G)$.
If $\hat{d}(G') = \hat{d}(G)$, then
$\mu(G',k')+2\hat{d}(G') = \mu_{\hat{d}(G)}(G',k')+2\hat{d}(G)$.
Otherwise, $\hat{d}(G') \leq \hat{d}(G)-1$.
Therefore, by Property~(P\ref{prop-range}), 
$\mu(G',k')+2\hat{d}(G') \leq \mu_{\hat{d}(G)}(G',k')+2\hat{d}(G)$.
\end{proof}

\begin{lemma}\label{lem:leaves}
$L(G,k) \leq 2.0409^{\mu(G,k)+2\hat{d}(G)+\alpha(G,k)}$.
\end{lemma}
\begin{proof}
Let $b = 2.0409$.
We prove the lemma by induction on the height of the recursion tree.
The base of the induction holds since
$\mu(G,k)+2\hat{d}(G)+\alpha(G,k) \geq \mu(G,k)+6 \geq -\Psi_{\hat{d}(G)}(m_2(G),c_2(G))+6 > 0$
(the last inequality holds due to Property~(P\ref{prop-range})).

We now prove the induction step.
Suppose that the algorithm applies a reduction rule on $(G,k)$,
and let $(G',k')$ be the resulting instance.
We have that
\begin{align*}
L(G,k)
& = L(G',k')\\
& \leq b^{\mu(G',k')+2\hat{d}(G')+\alpha(G',k')}\\
& \leq b^{\mu_{\hat{d}(G)}(G',k')+2\hat{d}(G)+\alpha(G,k)}\\
& \leq b^{\mu_{\hat{d}(G)}(G,k)+2\hat{d}(G)+\alpha(G,k)}\\
& = b^{\mu(G,k)+2\hat{d}(G)+\alpha(G,k)}
\end{align*}
where the first inequality follows from the induction hypothesis,
the second inequality follows from Lemma~\ref{lem:mu} and the definition of $\alpha$,
and the last inequality  follows from Lemma~\ref{lem:reduction}.

Now suppose that Rule~(\ref{rule:not-connected}) is applied on $(G,k)$.
Let $k'$ be the minimum size of a hitting set of $H$.
Since Rule~(\ref{rule:cc}) was not applied, then $k' \geq 4$.
We have
\begin{align*}
L(G,k) &= L(G-H,k-k') + \sum_{i=4}^{\min(k',k-4)} L(H,i)\\
& \leq
b^{\mu(G-H,k-k')+2\hat{d}(G-H)+\alpha(G-H,k-k')} + \sum_{i=4}^{k-4} b^{\mu(H,i)+2\hat{d}(H)+\alpha(H,i)}\\
& \leq
b^{k-k'+2\hat{d}(G)+\alpha(G,k)} + \sum_{i=4}^{k-4} b^{i+2\hat{d}(G)+\alpha(G,k)}\\
& \leq
b^{k-4+2\hat{d}(G)+\alpha(G,k)} + b^{k-3+2\hat{d}(G)+\alpha(G,k)}\\
& \leq
b^{k-2+2\hat{d}(G)+\alpha(G,k)}\\
& \leq
b^{\mu(G,k)+2\hat{d}(G)+\alpha(G,k)},
\end{align*}
where the first inequality follows from the induction hypothesis,
the second inequality follows from Property~(P\ref{prop-range}),
the third inequality follows from the fact that $k' \geq 4$, and
the last inequality follows from Property~(P\ref{prop-range}).

Now suppose that Rule~(\ref{rule:final}) is applied on $(G,k)$ and $d(G) \leq 3$.
Let $(\instance{G}{+u},k-1)$ and $(\instance{G}{-u},k)$ be the resulting instances.
By definition, $\alpha(G,k) = 3$.
By Lemma~\ref{lem:path}, $\alpha(\instance{G}{+u},k-1) = 0$ and $\alpha(\instance{G}{-u},k) = 0$.
Therefore,
\begin{align*}
L(G,k) &= L(\instance{G}{+u},k-1)+L(\instance{G}{-u},k)\\
& \leq
b^{\mu(\instance{G}{+u},k-1)+2\hat{d}(\instance{G}{+u})} + b^{\mu(\instance{G}{-u},k)+2\hat{d}(\instance{G}{-u})}\\
& \leq
b^{k-1+2\hat{d}(G)} + b^{k+2\hat{d}(G)}\\
& \leq
b^{k+1+2\hat{d}(G)}\\
& \leq
b^{\mu(G,k)+2\hat{d}(G)+\alpha(G,k)},
\end{align*}
where the first inequality follows from the induction hypothesis,
the second inequality follows from Property~(P\ref{prop-range}),
and the last inequality follows from  Property~(P\ref{prop-range}).

Finally, suppose that a branching rule is applied on $(G,k)$ and the cases discussed above do not occur.
Let $(G_1,k_1),\ldots,(G_s,k_s)$ be the instances generated by the branching rule.
We have that 
\begin{align*}
L(G,k) & = \sum_{i=1}^s L(G_i,k_i)\\
& \leq
\sum_{i=1}^s b^{\mu(G_i,k_i)+2\hat{d}(G_i)+\alpha(G_i,k_i)}\\
& \leq
b^{\hat{d}(G)+\alpha(G,k)} \sum_{i=1}^s b^{\mu_{\hat{d}(G)}(G_i,k_i)}\\
& \leq
b^{\mu(G,k)+2\hat{d}(G)+\alpha(G,k)},
\end{align*}
where the first inequality follows from the induction hypothesis,
the second inequality follows from Lemma~\ref{lem:mu},
and the last inequality follows from the fact that the branching numbers of the branching rules
are at most $b$.
\end{proof}

We obtain the following theorem.
\begin{theorem}
There is an $O^*(2.0409^k)$-time algorithm for \textsc{$3$-Hitting Set}.
\end{theorem}
\begin{proof}
The correctness of algorithm \hsb\ follows from the safeness of its rules.
By Lemma~\ref{lem:leaves}, the running time of the algorithm is
$O^*(2.0409^{\mu(G,k)+2\hat{d}(G)+\alpha(G,k)}) = O^*(2.0409^k)$
(as $\mu(G,k) \leq k$, $\hat{d}(G) \leq 6$, and $\alpha(G,k) \leq 3$).
\end{proof}

\bibliographystyle{abbrv}
\bibliography{hitting}

\end{document}